\documentclass[twocolumn]{aastex631}

\usepackage{xspace}
\usepackage[flushleft]{threeparttable}
\usepackage{color}
\usepackage{url}
\usepackage{hyperref}
\usepackage{graphicx}

\newcommand {\flux}{erg\,s$^{-1}$\,cm$^{-2}$\xspace}
\newcommand {\lum}{erg\,s$^{-1}$\xspace}
\newcommand{\ms}{$\mathrm{M_\odot}$\xspace}
 
\newcommand {\nicer}{{NICER}\xspace} 
\newcommand {\ixpe}{{IXPE}\xspace}
\newcommand {\nustar}{{NuSTAR}\xspace}
\newcommand {\hxmt}{{Insight-HXMT}\xspace}
\newcommand {\sco}{\mbox{{Sco~X-1}}\xspace}

\shorttitle{X-ray polarimetry of Sco X-1}
\shortauthors{La Monaca et al.}

\begin{document}

\title{Highly Significant Detection of X-Ray Polarization from the Brightest Accreting Neutron Star Sco~X-1}

\correspondingauthor{Fabio La Monaca}
\email{fabio.lamonaca@inaf.it}
\author[0000-0001-8916-4156]{Fabio {La Monaca}}
\affiliation{INAF Istituto di Astrofisica e Planetologia Spaziali, Via del Fosso del Cavaliere 100, 00133 Roma, Italy}
\affiliation{Dipartimento di Fisica, Universit\`{a} degli Studi di Roma ``Tor Vergata'', Via della Ricerca Scientifica 1, 00133 Roma, Italy}
\affiliation{Dipartimento di Fisica, Universit\`{a} degli Studi di Roma ``La Sapienza'', Piazzale Aldo Moro 5, 00185 Roma, Italy}

\author[0000-0003-0331-3259]{Alessandro Di Marco}
\affiliation{INAF Istituto di Astrofisica e Planetologia Spaziali, Via del Fosso del Cavaliere 100, 00133 Roma, Italy}

\author[0000-0002-0983-0049]{Juri Poutanen}
\affiliation{Department of Physics and Astronomy,  20014 University of Turku, Finland}

\author[0000-0002-4576-9337]{Matteo Bachetti}
\affiliation{INAF Osservatorio Astronomico di Cagliari, Via della Scienza 5, 09047 Selargius (CA), Italy}
\author[0000-0002-6154-5843]{Sara Elisa Motta}
\affiliation{INAF Osservatorio Astronomico di Brera, Via E. Bianchi 46, 23807 Merate (LC), Italy}

\author[0000-0001-6289-7413]{Alessandro Papitto}
\affiliation{INAF Osservatorio Astronomico di Roma, Via Frascati 33, 00040 Monte Porzio Catone (RM), Italy}
\author[0000-0001-7397-8091]{Maura Pilia}
\affiliation{INAF Osservatorio Astronomico di Cagliari, Via della Scienza 5, 09047 Selargius (CA), Italy}
\author[0000-0002-0105-5826]{Fei Xie}
\affiliation{Guangxi Key Laboratory for Relativistic Astrophysics, School of Physical Science and Technology, Guangxi University, Nanning 530004, China}
\affiliation{INAF Istituto di Astrofisica e Planetologia Spaziali, Via del Fosso del Cavaliere 100, 00133 Roma, Italy}
\author[0000-0002-4622-4240]{Stefano Bianchi}
\affiliation{Dipartimento di Matematica e Fisica, Universit\`{a} degli Studi Roma Tre, Via della Vasca Navale 84, 00146 Roma, Italy}
\author[0009-0009-3183-9742]{Anna Bobrikova}
\affiliation{Department of Physics and Astronomy,  20014 University of Turku, Finland}
% ...

\collaboration{119}{(IXPE Collaboration)}



\begin{abstract}
The {Imaging X-ray Polarimetry Explorer} (IXPE) measured with high significance the X-ray polarization of the brightest Z-source Scorpius~X-1, resulting in the nominal 2--8\,keV energy band in a polarization degree of 1.0\% $\pm$ 0.2\% and a polarization angle of $8\degr \pm 6\degr$ at 90\% of confidence level. This observation was strictly simultaneous with observations performed by \nicer, \nustar, and \hxmt, which allowed for a precise characterization of its broad-band spectrum from soft to hard X-rays. The source has been observed mainly in its soft state, with short periods of flaring. We also observed low-frequency quasi-periodic oscillations. From a spectro-polarimetric analysis, we associate a polarization to the accretion disk at $<3.2$\% at 90\% of confidence level, compatible with expectations for an electron-scattering dominated optically thick atmosphere at the \sco inclination of $\sim$44\degr; for the higher-energy Comptonized component, we obtain a polarization of $1.3\%\pm0.4\%$, in agreement with expectations for a slab of Thomson optical depth of $\sim$7 and an electron temperature of $\sim$3~keV. 
A polarization rotation with respect to previous observations by OSO-8 and PolarLight, and also with respect to the radio-jet position angle, is observed. This result may indicate a variation of the polarization with the source state that can be related to relativistic precession or to a change in the corona geometry with the accretion flow.
\end{abstract}


\section{Introduction} \label{sec:intro}

Accreting weakly magnetized neutron stars (NSs) in low-mass X-ray binaries (LMXBs) accrete matter via Roche lobe overflow from a stellar companion of mass smaller than one solar mass \citep{Bahramian2022}. The result of the matter overflowing the Roche lobe is the formation of an accretion disk around the compact object.
Contrary to the case of black holes (BHs), the accreting gas interacts with the NS surface forming a boundary layer (BL) coplanar to the disk \citep{Shakura88,Popham01}. However, the gas is not stopped immediately but spreads along the NS surface to high latitudes forming a spreading layer \citep[SL;][]{inogamov1999,suleimanov2006}. 

NS-LMXBs are very bright X-ray sources, and by studying their timing properties in the 1--10\,keV band, and/or their tracks in the X-ray hard-color/soft-color diagram (CCD), or their hard-color/intensity diagram (HID), they can be classified as Z- and atoll sources \citep{vanderklis89,hasinger89}.
The luminosity of Z-sources is higher ($\gtrsim10^{37}$\,\lum, near or just below the  Eddington luminosity) than that of the atoll sources ($10^{36}$--$10^{37}$\,\lum), and they show a Z-like track in the CCD. Along such a track, three different branches can be clearly distinguished: the horizontal (HB), the normal (NB), and the flaring branch (FB). They are correlated with the mass accretion rate, and not all these branches are well distinguished in every Z-source \citep{Schulz93,Kuulkers94}. In addition, we can also identify a hard apex (HA), that separates HB and NB, and a soft apex (SA) that separates NB and FB \citep{Church12, Motta19}. Z-sources can also be classified as Cyg-like and Sco-like sources \citep{Kuulkers94, kuulkers97}. \mbox{Cyg X-2}, \mbox{GX 5$-$1}, and \mbox{GX 340+0} belong to the first group, while \sco, \mbox{GX 349+2}, and \mbox{GX 17+2} to the latter one. Cyg-like sources show the full Z pattern with all the branches well visible; however, they have weak flaring compared to Sco-like sources. On the contrary, most of the latter ones show an HB that is hard to identify, and the flaring is more frequent and with a higher X-ray flux \citep{Church12}. Moreover, Cyg-like sources may have a higher inclination \citep{kuulkers97} or a higher magnetic field with respect to the Sco-like ones \citep{Psaltis95}.

The X-ray spectrum of weakly magnetized NS-LMXBs can be typically described by two main components. The first is a blackbody or multicolor disk, which describes the softer thermal spectral emission from the accretion disk and/or the NS surface; the second is a harder Comptonization component, which describes the emission from electrons accelerated at the boundary between the accretion disk and the NS surface that occurs in the BL/SL region. In addition to these soft and hard components, an iron emission line at $\sim$\,6.4\,keV can be observed in several weakly magnetized NSs as the prominent feature of reflection from the inner accretion disk. This is clear evidence of the presence of an external source of hard X-rays with respect to the colder disk, where the reflection process takes place. 
The iron line broadening is shaped by the gravitational redshift, the Doppler shift, and the relativistic Doppler boosting, giving a direct way to probe the physics of the region of strong gravity close to the compact object \citep[see, e.g.,][]{Cackett08, Cackett10, Cackett12, Matt06, Ludlam18, Ludlam22, Mondal20, Mondal22}. In particular, the reflection component allows to determine the inner radius of the accretion disk and, applying suitable models, can provide estimates for the proximity of the illumination source to the disk through the emissivity profile and the reflection fraction \citep[see, e.g.,][]{Wilkins2018}.
The spectral and timing properties of these sources allow us to understand the basics of their emission mechanisms, but they are not able to fully constrain the geometry of the different emitting regions, because the spectral features are degenerate from this point of view. At present, observations favor a scenario in which the BL/SL spectrum, represented to a certain accuracy by the Fourier frequency resolved spectrum, remains constant in the course of the luminosity variations
\citep{gilfanov2003,Revnivtsev06,Revnivtsev13}. This also means that Quasi-Periodic Oscillations (QPOs) are associated with the harder spectral component, which is due to the contribution of BL/SL.

A tool to disentangle the possible emission geometries is X-ray polarization. In fact, different geometries of the corona can be associated with different X-ray polarization properties \citep[as in the case of BH-LMXBs,][]{Haardt93,Poutanen93,PS96,Schnittman2010}, or by the accretion disk \citep{Chandrasekhar1960,Loskutov82}, or by the BL/SL \citep{Lapidus85,suleimanov2006}.
Almost 40 years after the first tentative attempts to measure X-ray polarization, the recent launch of the Imaging X-ray Polarimetry Explorer  \citep[\ixpe,][]{Weisskopf2022,Soffitta21}  adds  X-ray polarization to spectral and timing information to investigate the geometry of the accretion flow in LMXBs. Aiming to obtain this, \ixpe observed several weakly-magnetized LMXBs: the two Z-sources \mbox{Cyg X-2} \citep{Farinelli23} and \mbox{GX 5$-$1} \citep{Fabiani23}, the Z-atoll transient  \mbox{XTE J1701$-$462} \citep{Jayasurya2023, Cocchi2023}, the peculiar source \mbox{Cir X-1} \citep{Rankin23}, and three atoll sources: \mbox{GS~1826$-$238} \citep{Capitanio23}, \mbox{GX~9+9} \citep{Chatterjee2023, Ursini23} and the ultra compact \mbox{4U~1820$-$303} \citep{DiMarco23}. The average 2--8\,keV polarization is higher in the Z-sources than in the atoll ones, and it seems to vary with the position of the sources on the Z-track in the CCD \citep{Cocchi2023, Fabiani23, Rankin23}. For the Z-source Cyg X-2, \ixpe measured a polarization angle (PA) that appears to be aligned with the radio-jet \citep{Farinelli23}. 

\sco is a weakly magnetized NS-LMXB, and it is the brightest persistent and the first discovered extrasolar X-ray source \citep{Giacconi62}. It has a peak luminosity near the Eddington limit, that is, for a 1.4\,$\mathrm{M_\odot}$ NS, $\sim$2$\times10^{38}$\,\lum 
\citep{Titarchuk14}. 
The companion is an M-type star of $\sim$0.4\,\ms \citep{Steeghs02}. The orbital period of the system estimated from optical observations is $\sim$19\,hr \citep{Gottlieb75, Galloway14}.  The distance to Sco X-1, $D= 2.13^{+0.21}_{-0.26}$\,kpc, as derived in GAIA Data Release 2 \citep{Arnason21}, is in agreement with the previous estimates obtained by parallax measured with Very Long Baseline Array \citep{Bradshaw99}. Sco X-1 was the first X-ray binary where radio emission was also detected \citep{Andrew68}. 
VLBI observations spatially resolved the jet at sub-milliarcsecond scales revealing mildly relativistic motion of components in opposite directions at a position angle of $\sim$54\degr\ and inclination $44\degr \pm 6\degr$ \citep{Fomalont01,Fomalont2001b}.

Being the brightest X-ray source in the sky, apart from the Sun, \sco was one of the first targets for  X-ray polarization studies. 
In 1977, the OSO-8 satellite spent about 15~days observing it, obtaining a low-significance detection at 2.6\,keV with the polarization degree (PD) of $0.4\%\pm0.2\%$ and PA of $29\degr\pm10\degr$, and at 5.2\,keV PD=$1.3\%\pm0.4\%$ and PA=$57\degr\pm6\degr$ \citep{Long1979}. Recently, PolarLight observed \sco for a total of $\sim$322~days \citep{Long2022}, obtaining hints for variations of polarization with the energy and the source flux. In this case, a 5$\sigma$ detection in the 4--8\,keV energy band was obtained, but only when the flux is high: PD=$4.3\%\pm0.8\%$, and PA=$53\degr \pm 5\degr$. Although the polarization properties change with the emission state \citep{Cocchi2023,Fabiani23,Rankin23}, these measurements were performed in long exposures for which the emission state could not be determined, thus limiting the usefulness of these results.
\begin{figure*}
\centering   
\includegraphics[width=0.95\linewidth]{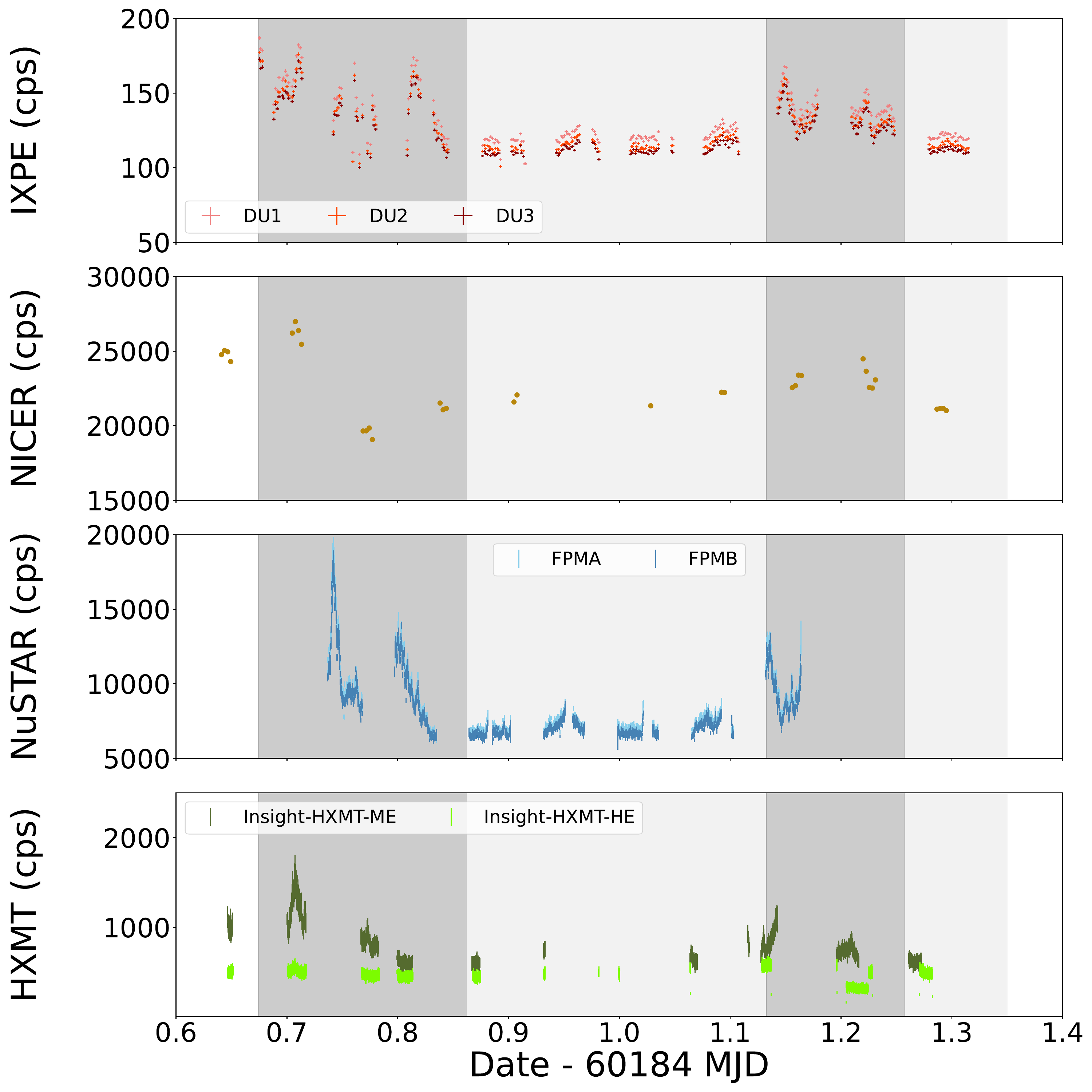}
\caption{Light curves from the simultaneous observations of \sco, from top to bottom: \ixpe (DU1, DU2, and DU3), \nicer, \nustar (FPMA, and FPMB), and \hxmt. They show that all the observations are well overlapped with the IXPE one. The flaring activity of the source is seen in all the simultaneous observations (dark gray shaded regions), while the non-flaring periods are identified with the light gray bands. 
}
\label{fig:sco_lc_all}
\end{figure*}

\begin{deluxetable}{llcc} 
\tablecaption{Exposure time in each observatory, for each telescope/instrument on board and observation ID.}
\label{tab:exposure}
\tablehead{&\colhead{Obs ID} & \colhead{Telescope} &\colhead{Exp. time (s)} 
 }
\startdata
\ixpe & 02002401 & DU 1 & 24456 \\
& & DU 2 & 24456 \\
& & DU 3 & 24456 \\
\hline
\nicer & 6689020201 & & 3039 \\
& 6689010102 & & 2537 \\
\hline
\nustar & 30902036002 & FPMA & 932 \\
& & FPMB & 1011 \\
\hline
HXMT & P060531600101 & ME & 2093 \\
& & HE & 3210 \\
& P060531600102 & ME & 1236 \\
& & HE & 1839 \\
& P060531600103 & ME & 80 \\
& & HE & 235 \\
& P060531600104 & ME & 1324\\
& & HE & 687\\
& P060531600105 & ME & 1796\\
& & HE & 2460 \\
\enddata
\end{deluxetable}

\section{X-ray Observations} \label{sec:Observations}

\ixpe observed \sco on 2023 August 28 starting at 16:10  UTC (ObsID: 02002401) for a total exposure of $\simeq$24\,ks per Detector Unit (DU). 
Because of the high X-ray flux from \sco, the \ixpe observations were performed using a partially opaque absorber, the so-called gray filter present in the detector's filter and calibration wheel \citep{Ferrazzoli20, Soffitta21, Weisskopf2022}. For such a bright source, this filter is needed to reduce the flux to a level compatible with the \ixpe focal plane detector dead-time. In particular, it allows for a tenfold reduction of the incident flux at low energies, below 3 keV. For the data analysis, we applied the most recent response matrices released in \textsc{HEASoft} on 2023 June 16. 

\begin{figure}
\centering
\includegraphics[width=1.0\linewidth]{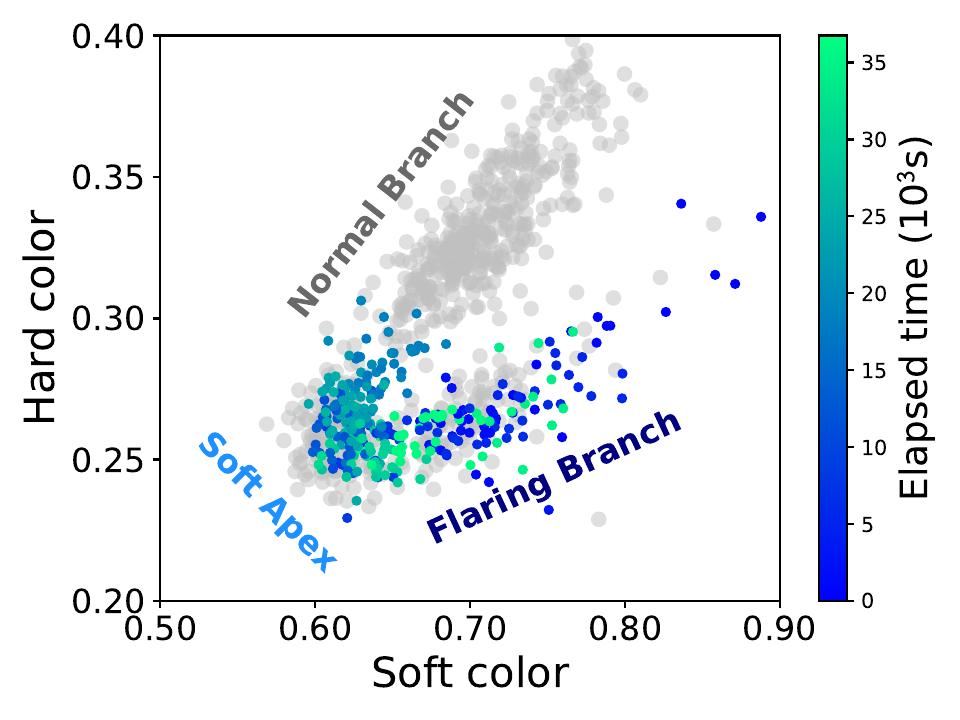}
\caption{Color-color diagram from the \nustar data. Soft color is defined as the source count rate in the 6--10\,keV band divided by the one in 3--6\,keV, while the hard color is the ratio of the 10--20\,keV count rate to the one in 6--10\,keV. Gray points represent the past \nustar observations, while the colored ones report from blue to green the elapsed time since the start of the present \nustar observation simultaneous with \ixpe. 
}
\label{fig:CCD}
\end{figure}

\begin{figure} 
\centering
\includegraphics[width=0.95\linewidth]{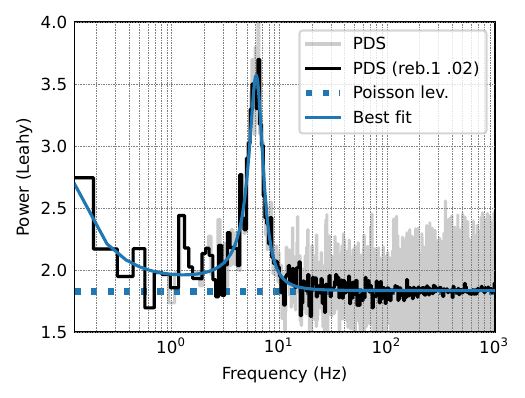}
\caption{\citet{bartlett1948}-like periodogram from the last $805$ seconds of \nicer ObsID 6689010102, obtained by averaging 100 periodograms from 8-sec segments of the light curve, normalized as in \citet{Leahy83} (gray), and re-binned geometrically (factor 1.02) to improve signal-to-noise at high frequencies (black). In blue, we plot the best-fit model consisting of a Poisson noise level, significantly below 2 due to dead time, a power-law red-noise component at low frequencies, and a QPO.}
\label{fig:pds}
\end{figure}

Jointly with \ixpe, we planned strictly simultaneous observations with other facilities to have a better energy resolution and to cover a broader energy band than those available with \ixpe alone, aiming to obtain an accurate spectral model of the source simultaneous to the measurement of polarization. Our coordinated observations were: i) \nustar (ObsID 30902336002) from 2023 August 28 at 17:41 UTC to August 29 at 04:41 UTC; ii) \nicer (ObsIDs 6689010101, 6689010102) on 2023 August 28 from 15:23 to 23:14 UTC and on 2023 August 29 from 00:40 to 07:07 UTC, 
iii) \hxmt (ObsIDs P06053160010N, with N=1--5), from 2023 August 28 at 15:30 up to 19:11, from 2023 August 28 at 19:11 up to 22:21, from 2023 August 28 at 22:21 up to 2023 August 29 at 01:31, from 2023 August 29 at 1:31 up to 4:42, and from 2023 August 29 at 4:42 up to 7:27. The exposure times, after the dead-time correction, in each ObsID, and for each instrument are reported in Table \ref{tab:exposure}. 
For \hxmt, the Low Energy instrument is not well calibrated, and we excluded it from the analysis; the Medium (ME) and High Energy (HE) instruments were used in the analysis. 

In Figure~\ref{fig:sco_lc_all}, we show the light curves of the source as observed with all the coordinated X-ray observatories. We note that all of them were strictly simultaneous with \ixpe.
We also observe a flaring activity, causing a simultaneous increase in the count rate of the source for all the telescopes covering different energy bands.

To investigate the state of the source during the observations, we calculated the CCD considering all the observations of \sco in the \nustar data archive (ObsIDs 30001040002 and 30502010002). In Figure~\ref{fig:CCD}, we show the CCD obtained from the archival observations together with those performed during the \ixpe run. 
We can conclude that during the IXPE observation, \sco is moving from the FB to the NB and then again to the FB of the Z track. The two passages and the fact that the source never entered deeply into the NB, produced an IXPE observation dominated by the SA, with short periods in the FB.

We detected a transient QPO in \nicer ObsID 6689010102. Using the \textsc{stingray} package \citep{huppenkothen2019}, we produced a \citet{bartlett1948} periodogram (an estimate of the power density spectrum, PDS) of the full obsID, averaging the periodograms calculated in 153 intervals of equal length (16\,s each).
We found a clear $\sim$6\,Hz QPO, and we proceeded to verify whether its properties changed during the observation. We then inspected a dynamical power spectrum, showing the change of the QPO in subsets of the observation, and it was immediately clear that the QPO disappeared after the first good time interval (GTI) of the observation and reappeared in the last, with no major changes of amplitude or frequency inside a given GTI. We proceeded to analyze the periodograms on a per-GTI basis. This time, we chose a smaller segment size (8\,s) to improve the statistics at the expense of some frequency resolution. We fit a model comprising a Poisson noise level, a QPO at $\sim$6\,Hz, and an optional power law when required by the low-frequency data. We used the maximum-likelihood method, based on the Whittle statistics, described by \citet{barret2012} and implemented in the \textsc{stingray.modeling} subpackage.
Since the analysis of the QPO is not the main focus of the paper, we quote the $1\sigma$ error bars from the fitting routine used internally by \textsc{stingray}, knowing that the estimates might be slightly off. In the first GTI (31\,s), the QPO had a central frequency of 7.2$\pm$0.2\,Hz peak and a FWHM of 3.0$\pm$0.5\,Hz (quality factor $\nu/\Delta\nu\approx2.4$) at 68\% of confidence level (CL). 
In the last GTI (805\,s), the peak was at 6.07$\pm$0.01\,Hz, and the FWHM was 2.27$\pm$0.01\,Hz ($\nu/\Delta\nu\approx2.7$) at 68\% CL. Figure~\ref{fig:pds} shows the second QPO and the best-fit model. These frequencies are compatible with the CCD result: the source, in the beginning, was in a flaring state (QPO$\sim$7 Hz), then moved toward the NB (but never entered it), staying in the SA, just before the end of the observation a short flaring state is observed, then the source came back to the SA, where a 6--7 Hz range for QPOs is expected as reported, e.g., by \citet{Casella06}.

\begin{deluxetable*}{ccccccccccc}
\tablewidth{0pt}
\tablecaption{Polarization properties of \sco obtained with the \texttt{PCUBE} algorithm and with the spectro-polarimetric \textsc{xspec} analysis, applied to Model A of Table~\ref{tab:spectrum}.}
\label{tab:pol_pcube_xspec}
\tablehead{
&\multicolumn{4}{c}{\texttt{PCUBE}} & &
\multicolumn{4}{c}{\textsc{xspec}} \\ 
\cline{2-5} \cline{7-10}
\colhead{Energy bin} & \colhead{PD} & \colhead{PA} & 
\colhead{$Q/I$} & \colhead{$U/I$}  & &\colhead{PD} & \colhead{PA} & 
\colhead{$Q/I$} & \colhead{$U/I$} \\
\colhead{(keV)}  & \colhead{(\%)} & \colhead{(deg)} & \colhead{(\%)} & \colhead{(\%)} & & \colhead{(\%)} &  \colhead{(deg)} & \colhead{(\%)} & \colhead{(\%)}}     
\startdata
2--3    & --&--&--&--&& $0.9\pm0.3$ &  $-4\pm10$ & $0.9\pm0.3$ & $-0.1\pm0.3$\\
3--4    & $0.9\pm0.2$ & $7\pm7$ & $0.9\pm0.2$ & $0.2\pm0.2$          && $0.9\pm0.2$ & $7\pm7$ & $0.9\pm0.2$                        & $0.2\pm0.2$\\ 
4--5    & $1.3\pm0.3$ & $12\pm6$ & $1.2\pm0.3$ & $0.5\pm0.3$         && $1.2\pm0.3$ & $12\pm6$ & $1.1\pm0.3$ 
        & $0.5\pm0.3$\\
5--6    & $1.5\pm0.3$ & $12\pm7$ & $1.4\pm0.3$ & $0.6\pm0.3$         && $1.5\pm0.3$ & $12\pm7$ & $1.3\pm0.3$                       & $0.6\pm0.3$\\ 
6--7    & $1.1\pm0.5$ & $8\pm13$ & $1.0\pm0.5$ & $0.3\pm0.5$         & & $1.1\pm0.5$ & $5\pm13$ & $1.1\pm0.5$                       & $0.2\pm0.5$\\
7--8    & $1.0\pm0.9$ & $22\pm26$ & $0.7\pm0.9$ & $0.7\pm0.9$ & &$0.7^{+0.9}_{-0.7}$ &$14\pm14$ & $0.6\pm0.9$ & $0.3\pm0.9$\\ 
\hline
3--8    & $1.08\pm0.15$ & $10\pm4$ & $1.02\pm0.15$ & $0.38\pm0.15$ && $1.08\pm0.15$ & $10\pm4$ & $1.01\pm0.14$ & $0.37\pm0.14$ \\ 
4--8    & $1.25\pm0.20$ & $12\pm5$ & $1.11\pm0.20$ & $0.51\pm0.20$ && $1.25\pm0.18$ & $11\pm4$ & $1.16\pm0.18$ & $0.48\pm0.18$
\enddata
\tablecomments{Errors are reported at 68\% CL.}
\end{deluxetable*}
\section{Polarimetric analysis}

We estimated the X-ray polarization by using \textsc{ixpeobssim} software \citep{Baldini22} with the \texttt{PCUBE} algorithm based on the \citet{Kislat2015} method, which allows for a model-independent analysis of the polarization. 

As reported in the \textsc{ixpeobssim} notes\footnote{\href{https://github.com/lucabaldini/ixpeobssim/issues/714}{https://github.com/lucabaldini/ixpeobssim/issues/714}} and in \citet{Veledina23}, the polarization obtained by the \texttt{PCUBE} algorithm, 
when using the gray filter, does not properly account for the telescope response matrices at low energy. In particular, an overestimate of the polarization below 3\,keV is observed with respect to the one obtained by the more solid \textsc{xspec} \citep{Arnaud96} spectro-polarimetric analysis, which properly considers the response matrices of the instrument at low energy. The spectro-polarimetric analysis is reported in detail in Section~\ref{sec:spectro-pol}.
\begin{figure} 
\centering
\includegraphics[width=0.8\linewidth]{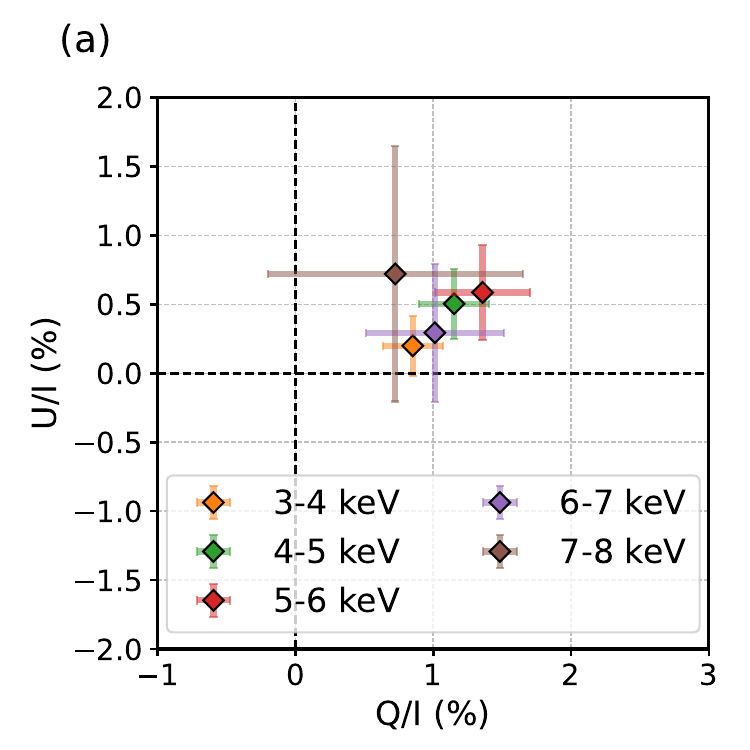} \\
\includegraphics[width=0.8\linewidth]{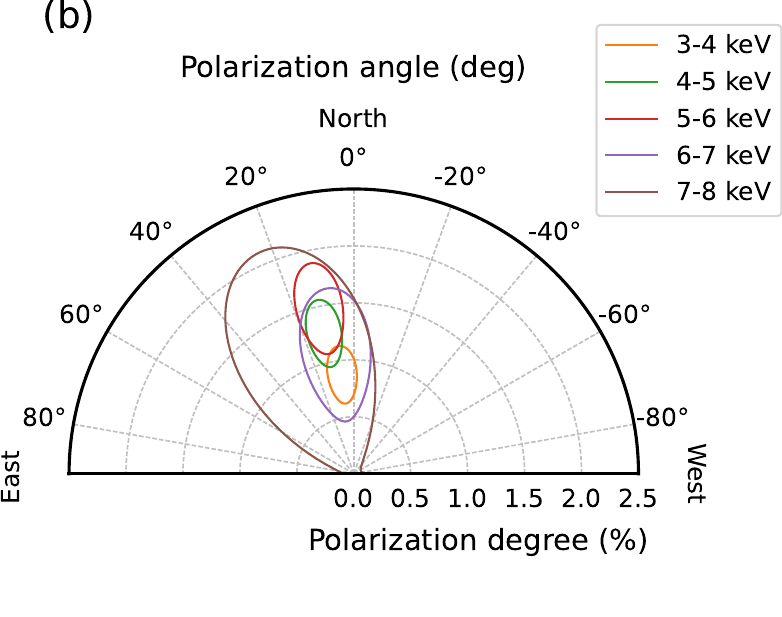} \\
\includegraphics[width=0.8\linewidth]{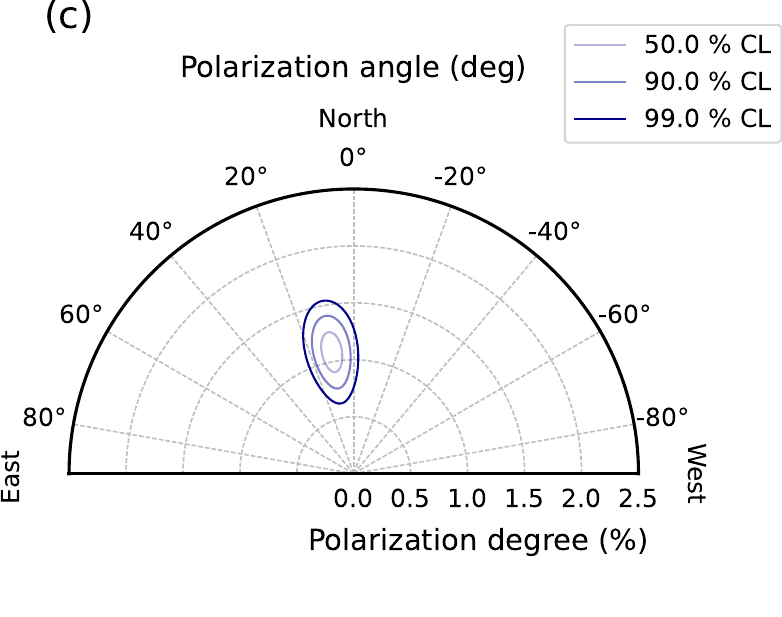}
\caption{In these plots the \sco polarization in different energy bins, as measured with \texttt{PCUBE} in \textsc{ixpeobssim}, are reported. (a) Normalized Stokes parameters plane in 1\,keV energy bins: $Q/I$ vs. $U/I$. (b) Polar plot of the X-ray polarization when 1 keV binning is applied; contours are given at 68\% CL. No trend of polarization varying with energy is visible neither in (a), and (b). (c) Polar plot of the X-ray polarization in the 3--8\,keV energy band; contours correspond to 50\%, 95\%, and 99\% CL. The detection significance of the energy averaged polarization is at $\sim$7$\sigma$ CL.
}
\label{fig:stokes_pcube}
\end{figure}
Therefore, we decided to use the \texttt{PCUBE} algorithm only in the 3--8\,keV band. We analyzed the \ixpe data, following the prescription reported in \citet{Di_Marco2023}. For bright sources, the background region is dominated by the source counting rate itself, that is any way at least one order of magnitude lower than the source signal on the whole \ixpe energy band. Therefore, no background rejection is needed, and no background subtraction has to be performed. 

We analyzed the polarization as a function of the energy with bins of 1\,keV, results are shown in Figure~\ref{fig:stokes_pcube}a and reported in Table~\ref{tab:pol_pcube_xspec}. The Stokes parameters show no evidence of a PD variation with energy or any rotation of the PA, as can also be seen in the polar plot for the X-ray polarization of Figure~\ref{fig:stokes_pcube}b, where the contours at 68\% CL for all the energy bins are well compatible. Thus, we can combine all the energy bins to derive the polarization in the 3--8\,keV energy band, as reported both in Figures~\ref{fig:stokes_pcube}c and Table~\ref{tab:pol_pcube_xspec}. The resulting polarization is PD$=1.08\%\pm0.15\%$, and PA$=10\degr\pm4\degr$ at 68\% CL. The probability of obtaining such polarization in the case of an unpolarized source is $6\times 10^{-12}$, corresponding to a detection significant at $\sim$7$\sigma$ CL.

\begin{figure}[!tb]
\centering\includegraphics[width=0.95\linewidth]{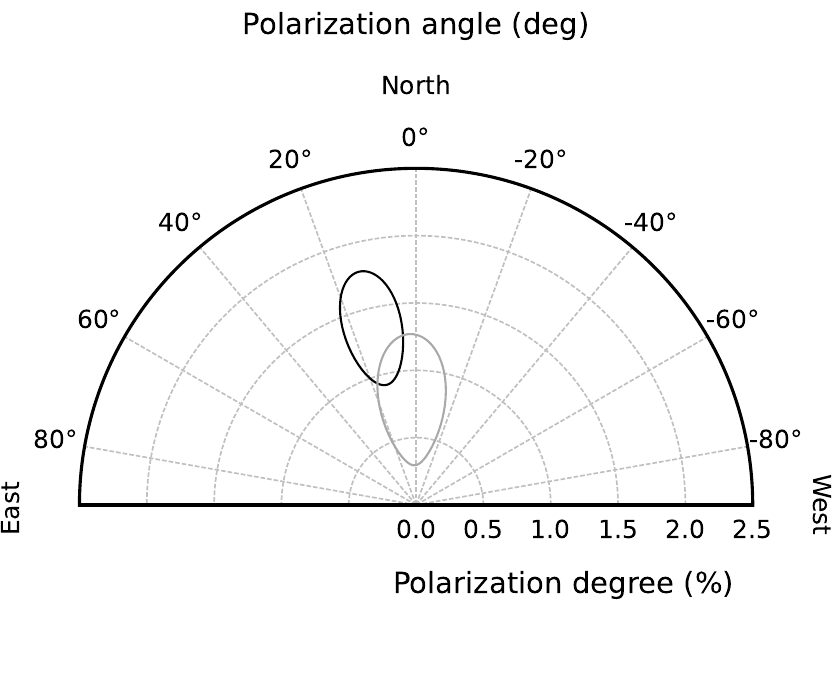} 
\caption{Polar plot of the X-ray polarization for \sco in the two different flaring (dark gray) and non-flaring (light gray) states; the two states show a compatible polarization. The contours are at 90\% CL.}
\label{fig:contour_flaring}
\end{figure}

We also tried to separate the flaring activity of the source from the non-flaring state. The two different time intervals are defined by the shaded regions reported in Figure~\ref{fig:sco_lc_all}. The polarimetric results are reported in Figure~\ref{fig:contour_flaring}, which shows that the confidence regions for the two intervals are compatible within 90\% CL, although it seems to present a marginal indication for a slightly higher PD in the flaring state. Therefore, considering that present data do not allow for clearly observing a PD variation in the two states, and, as reported in \citet{Mazzola21}, that no strong spectral variation in the different \sco states is reported, the analysis that follows will be carried out without dividing into flaring and non-flaring states.

\begin{deluxetable*}{crcccc}
\tablecaption{Best-fit parameters of the spectral models applied to the simultaneous data from NICER, \nustar, \hxmt, and \ixpe. }
\label{tab:spectrum}
\tablehead{
\colhead{Component} &
\colhead{Parameter (units)} &
\colhead{Continuum} &
\colhead{Model A} &
\colhead{Model B} &
\colhead{Model C} }
\startdata
\texttt{tbabs} & $N_{\rm H}$ ($10^{22}$ cm$^{-2}$) & $0.15$ & $0.15$ & $0.15$ & $0.15$ \\
\hline
\texttt{diskbb} & $kT_{\rm in}$ (keV) 
& $0.692\pm0.004$ & $0.683\pm0.011$ & $0.694\pm0.01$ & $0.87\pm0.02$\\
& \text{norm} ($[R_{\rm in}/D_{10}]^2\cos\theta$) 
& $25500\pm700$ & $25500\pm1400$ & $24300\pm1200$ & $10800\pm800$\\
& $R_{\rm in}$\tablenotemark{a} (km) & 
$40.1\pm1.1$ & $40\pm2$ & $39\pm2$ & $26\pm2$ \\
\hline
\texttt{nthcomp} & $\Gamma$ & 
$2.78\pm0.02$ & $2.58\pm0.02$ & $2.61\pm0.03$ & $2.59\pm0.03$ \\
& $\tau$ & $5.95\pm0.04$ & $6.70\pm0.05$ & $6.62\pm0.08$ & $6.71\pm0.08$\\
& $kT_{\rm e}$ (keV) & 
$3.41\pm0.03$ & $3.21\pm0.03$ & $3.23\pm0.03$ & $3.22\pm0.04$ \\
& $kT_{\rm bb}$ (keV) & 
$1.148\pm0.008$ & $1.08\pm0.01$ & $1.10\pm0.01$ & $1.24\pm0.02$\\
& \text{norm} &
$6.34 \pm 0.05$ & $6.94 \pm 0.16$ & $6.74 \pm 0.17$ & $4.19 \pm 0.17$ \\
\hline
\texttt{gauss} & $E_{\rm line}$ (keV) &
-- & $6.72\pm0.02$ & -- & -- \\
& $\sigma$ (keV) & 
-- & $0.45\pm0.03$ & -- & --\\
& \text{norm} (photon~cm$^{-2}$~s$^{-1}$) &
-- & $0.14\pm0.01$ & -- & -- \\
& Equivalent width (eV) &
-- & 66.836$\pm$0.014 & -- & --\\ 
\hline
\texttt{diskline} & $E_{\rm line}$ (keV) &
-- & -- & $6.70\pm0.02$ & -- \\
& Emissivity &
-- & -- & $2.13\pm 0.09$ & --\\
& $R_{\rm in}$ ($GM/c^2$) &
-- & -- & $11\pm4$& --\\
& $R_{\rm out}$ ($GM/c^2$) &
-- & -- & 1000\tablenotemark{b} & --\\
& Inclination (deg) &
-- & -- & 44\tablenotemark{b} & --\\
& \text{norm} (photon~cm$^{-2}$~s$^{-1}$) &
-- & -- & $0.130^{+0.007}_{-0.003}$ & --\\
& Equivalent width (eV) &
-- & -- & 65.38$\pm$0.01 & --\\ 
\hline
\texttt{relxillNS} & Emissivity &
-- & -- & -- & $1.9\pm0.1$ \\
& $R_{\rm in}$ ($GM/c^2$)  &
-- & -- & -- & $<9$ \\
& $R_{\rm out}$ ($GM/c^2$) &
-- & -- & -- & 1000\tablenotemark{b}\\
& Inclination (deg) &
-- & -- & -- & 44\tablenotemark{b} \\
& $\log \xi$ &
-- & -- & -- & $2.48^{+0.04}_{-0.03}$ \\
& $A_{\rm Fe}$ &
-- & -- & -- & $4$\tablenotemark{b} \\
& $\log N$ &
-- & -- & -- & $19$\tablenotemark{b} \\
& \text{norm} &
-- & -- & -- & $0.078\pm0.004$\\
\hline
\multicolumn{2}{r}{$\chi^2$/dof} & 5365/2890 = 1.9 & 3033/2887 = 1.05 & 3083/2886 = 1.06 & 2988/2886 = 1.03\\
\hline
\multicolumn{6}{c}{Cross normalization factors} \\
   & $C_{\rm NICER}$ &
   1.0 & 1.0 & 1.0 & 1.0 \\
   & $C_{\rm \nustar-A}$ &
   $1.226\pm0.002$ & $1.226\pm0.002$ & $1.226\pm0.002$ & $1.225\pm0.002$\\
   & $C_{\rm \nustar-B}$ &
   $1.190\pm0.002$ & $1.190\pm0.002$ & $1.190\pm0.002$ & $1.190\pm0.002$\\
   & $C_{\rm HXMT-ME}$ &
   $1.144\pm0.003$ & $1.1462\pm0.003$ & $1.1463\pm0.003$ & $1.148\pm0.003$\\
   & $C_{\rm HXMT-HE}$ &
   $1.3\pm0.2$ & $1.4\pm0.2$ & $1.4\pm0.2$ & $1.4\pm0.2$\\
   & $C_{\rm \ixpe-DU1}$ &
   $1.090\pm0.002$ & $1.089\pm0.002$ & $1.089\pm0.002$ & $1.088\pm0.002$\\
   & $C_{\rm \ixpe-DU2}$ &
   $1.062\pm0.002$ & $1.061\pm0.002$ & $1.061\pm0.002$ & $1.061\pm0.002$\\
   & $C_{\rm \ixpe-DU3}$ &
   $1.004\pm0.002$ & $1.003\pm0.002$ & $1.003\pm0.002$ & $1.003\pm0.002$\\
    \hline
    \multicolumn{6}{c}{Photon flux ratios in 2--8\,keV} \\
   & $F_{\rm diskbb}/F_{\rm tot}$ & 0.23 & 0.21 & 0.22 & 0.32\\
  & $F_{\rm nthcomp}/F_{\rm tot}$ & 0.77 &  0.78 & 0.77 & 0.59\\
   & $F_{\rm gauss,diskline,relxillNS}/F_{\rm tot}$ & -- & 0.01 & 0.01 & 0.09\\
\enddata
\tablecomments{The \nicer flux in 2--8 keV is $1.98\times 10^{-7}$\flux, corresponding to the luminosity of 1.02$\times10^{38}$\,\lum ($\sim$0.6\,$L_{\rm Edd}$ for a 1.4\,$\mathrm{M_\odot}$ NS). 
Errors are reported at 90\% CL.}
\tablenotetext{a}{The inner radius for the \texttt{diskbb} component is estimated assuming an inclination at 44\degr, as reported in \citet{Fomalont01}, and a distance of 2.13\,kpc  \citep{Arnason21}.}
\tablenotetext{b}{Fixed.}
\end{deluxetable*}

\section{Spectral analysis}\label{sec:spectral_model}

\begin{figure}
\centering
\includegraphics[width=0.95\linewidth]{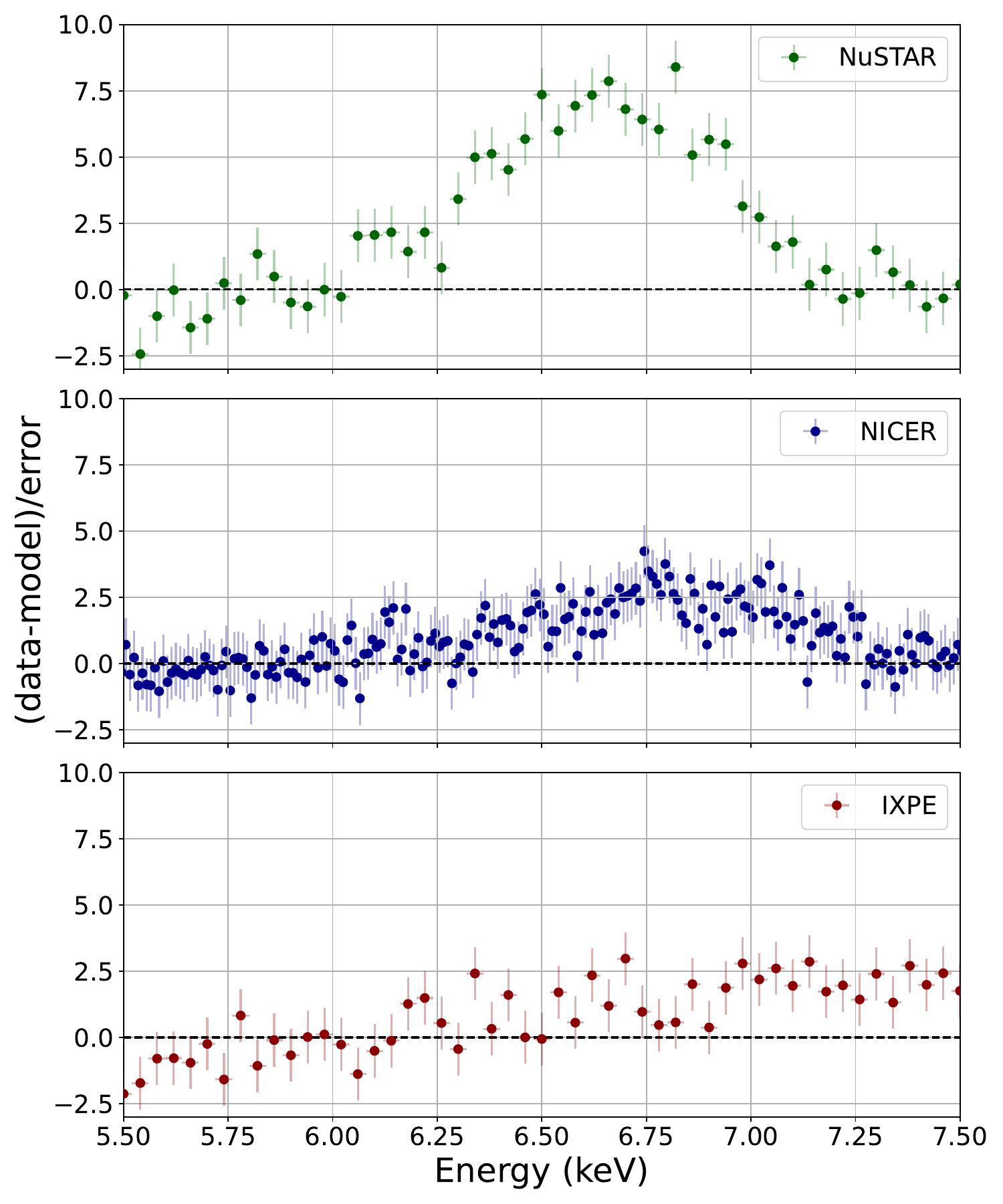}
\caption{Spectral residuals in the energy range 5.5--7.5\,keV for the \sco observations with \nustar-FPMA (top), NICER (center), and \ixpe-DU1 (bottom), when the spectral model which represents the continuum, without any component describing the Fe line complex. The \nustar and \nicer data show an excess compatible with a broad iron line, indistinguishable in the \ixpe spectrum due to its limited spectral capabilities.}
\label{fig:feline}
\end{figure}

Here, we performed spectral modeling for the coordinated observations using the data from: (i) \nicer in 1.5--12\,keV (we ignored data below 1.5\,keV, that can be affected by the Daytime Light Loading reported on May 2023); (ii) \nustar in 3--40\,keV; (iii) \hxmt Medium Energy (ME, in 8--30\,keV, ignoring 20--23\,keV energy band to avoid effects related to the presence of Ag in the Si-pin detectors \citep{Li20}) and High Energy (HE, in 30--40\,keV) telescopes; (iv) \ixpe in 2--8\,keV unweighted spectra \citep{DiMarco_2022}. In \nustar and \hxmt HE, events above 40\,keV are excluded in the analysis because the spectrum is dominated by background; see Appendix~\ref{sec:dataX} for details about data extraction. On the basis of the spectral models reported in the literature \citep{Mazzola21,Ding23}, we applied the simple model \texttt{tbabs*(diskbb+nthcomp)} in \textsc{xspec} v.12.13.0c \citep{Arnaud96} to describe the continuum. We also used energy-independent cross-normalization factors between the different X-ray observatories and telescopes. To estimate the absorption from the interstellar medium, we set the abundances at the \texttt{wilm} values \citep{2000ApJ...542..914W} and froze $N_{\rm H}$ at the galactic value of $0.15 \times 10^{22}$ cm$^{-2}$ \citep{Dickey90,HI4PI,Kalberla05}. About the \texttt{nthcomp} \citep{nthcomp1,nthcomp2}, we froze the \texttt{inp\_type} to 0, corresponding to blackbody seed photons, emitted e.g. from the neutron star SL/BL. The resulting fit for this model is reported in Table \ref{tab:spectrum}. 
We also estimated the optical depth $\tau$ of the Comptonization component from the asymptotic power-law photon index $\gamma$ of the \texttt{nthcomp}, using the relation:
\begin{eqnarray}\nonumber
\Gamma=\left[\frac{9}{4}+\frac{1}{\tau\left(1+\frac{\tau}{3}\right) \frac{kT_{\rm e}}{m_{\rm  e}c^2}}\right]^{1/2}-\frac{1}{2}\,, 
\label{eq:tau}
\end{eqnarray}
reported in \citet{nthcomp1}. The obtained values for the optical depth are reported in the same table. The continuum spectrum fit shows a bad $\chi^2$/dof (5365/2890=1.9) and a clear excess at $\sim$6.7\,keV due to the presence of an iron fluorescence line. This residual excess is shown in Figure~\ref{fig:feline}. Due to the worst spectral capabilities, this is not well visible in the \ixpe data but is evident in the \nustar and \nicer ones. Hereafter, no model-dependent systematic was applied to the fits, and when not explicitly reported, the uncertainties are always at 90\% CL. To take into account \nustar calibration uncertainties \citep{madsen22, Grefenstette22}, we allow the offset in the gain to vary by the quoted $\sim$40 eV systematic uncertainty and find a best-fit of $\sim$80\,eV for both the focal plane modules A and B (FPMA and FPMB); similarly for \ixpe calibration uncertainties \citep{DiMarco_2022b,Rankin23monitoring} we left free the gain slope and offset obtaining a slope of $\sim 0.94$\,keV$^{-1}$, and the offset $\sim 70$\,eV. Also, for \hxmt ME, we applied these gain corrections, obtaining $1.033\pm0.001$\,keV$^{-1}$ for the slope and an offset of $-0.205\pm0.014$ keV. No strong variations in the different adopted models of Table~\ref{tab:spectrum} are present, thus these values are representative of all of them. 

\begin{figure} 
\centering
\includegraphics[width=0.95\linewidth]{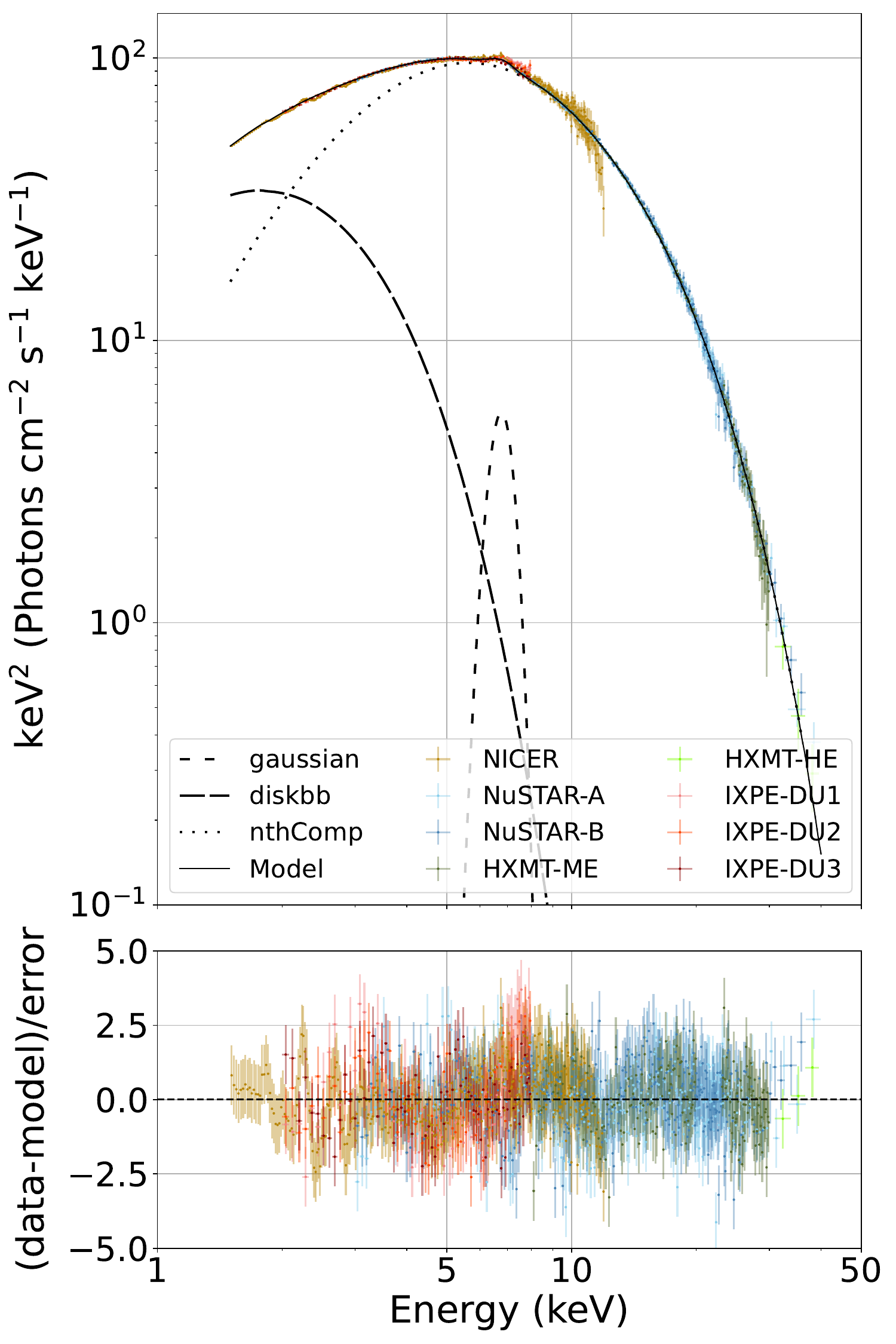}
\caption{Spectral energy distribution of \sco in $EF_E$ representation. The points show the data from NICER (brown), \nustar (blue), \hxmt (green), and \ixpe (red). The different spectral model components are reported in black lines for \texttt{diskbb} (dashed), \texttt{nthcomp} (dot-dashed), and \texttt{gauss} (dotted). 
The bottom panel shows the residuals between the  data and the best-fit Model A.}
\label{fig:bestfit}
\end{figure}

Aiming to describe the Fe line component, we tried to fit the spectrum with the Model A of Table~\ref{tab:spectrum}, which includes a broad Gaussian line, as in \citet{Di_Marco2023}. The resulting fit shows a  strong improvement of the $\chi^2$/dof (3033/2887 = 1.05) with reduction by a factor $\sim$1.8. The spectral fit for this model is shown in Figure~\ref{fig:bestfit}. 

Aiming to model this Fe line better, instead of a Gaussian, we tested a \texttt{diskline} component \citep{Fabian1989} in Model B: \texttt{tbabs*(diskbb+nthcomp+diskline)}. As reported in Table \ref{tab:spectrum}, the resulting $\chi^2$/dof is 3083/2886, with no improvement with respect to the Gaussian in terms of reduced $\chi^2$, but we obtain for the inner disk a radius of $11\pm4,R_{\rm g}$ \t 90\%CL, which corresponds to $1.83\,R_{\rm ISCO}$, meaning that the disk during the observation is near the NS surface. 

We have also attempted to fit with a complete reflection model, \texttt{relxillNS} in the \textsc{xillver} code \citep{Garcia22}, reported as Model C in Table \ref{tab:spectrum}: \texttt{tbabs*(diskbb+nthcomp+relxillNS)}. In this model, we assumed most of the X-ray continuum emission as thermal,  originating from either the surface of the NS or the SL/BL, and described by a blackbody with the same temperature as the seed photons in \texttt{nthcomp}. Then, it illuminates the accretion disk, producing the reprocessed/reflected X-ray spectrum of the reflection component. \texttt{xillverNS} \citep{Garcia22} solves this radiation transfer problem in a plane-parallel slab, providing an angle-dependent spectrum emergent at the top of this slab for a given irradiation, gas density, and elemental abundances. 
\texttt{relxillNS} used in this analysis represents a complete relativistic model taking into account the angular distribution of the solution provided by \texttt{xillverNS} and correctly predicts the integrated reflection of the disk by including all relativistic effects (such as distortion of the spectral features due to relativistic corrections such as boosting, gravitational redshift and Doppler effects). In this model, we froze the density $\log N$ parameter to 19, that is, the maximum value for which the model is calculated \citep{Garcia22}; the outer radius to $1000R_{\rm g}$ as in Model B; and the reflection fraction to $-1$ to obtain only the spectral component due to the reflected fraction, and the NS spin is fixed to O ($R_{\rm ISCO}=6R_{\rm g}$). The iron abundance, when it is left free to vary, tends to be as high as possible; thus, we fixed it to an acceptable value equal to 4, allowing us to still obtain a good reduced $\chi^2$ (this situation is similar to the case of Cyg X-1, where high iron abundance was inferred; \citealt{Tomsick18}). When left free to vary, the inclination becomes $79\degr\pm30\degr$ at 68\% CL, and it is unconstrained at 90\% CL. This value is compatible, given the large uncertainty, with the more accurate one from \citet{Fomalont01,Fomalont2001b}; for this reason, we froze it in the final analysis to the literature value of 44\degr.

When this model is applied, we obtain an improvement in terms of reduced $\chi^2$ and F-test (probability $7\times10^{-12}$) with respect to Model A. From this fit, we obtain an upper limit on the inner disk truncation radius of $9R_{\rm g}$, which corresponds to $1.50R_{\rm ISCO}$. This result is compatible with the one obtained by the \texttt{diskline} modeling, confirming the presence of a disk close to the NS surface. This model, when applied, shows a relative flux for the reflection component at a level of 10\%, which is a minor contribution with respect to the disk (26\% in this model) and the Comptonization, which is dominating in all the applied spectral models ($>64$\%).

The obtained results for the spectral model are consistent with the previous one reported by \citet{Mazzola21} for the continuum, while they do not apply a full reflection modeling, giving no result for the inner radius from this latter one. Therefore, in this paper, we obtained a measurement of the inner radius for \sco from the reflection modeling for the first time. This result is consistent with the one obtained for the Sco-like Z source GX 349+2 by \citet{349+2}, where an inner radius at the $R_{ISCO}$ level is found in the SA and in the FB states (except the extreme flaring one). A similar result was also obtained by \citet{Ludlam22} for Cyg X-2.

\begin{figure*} 
\centering
\includegraphics[width=0.82\textwidth]{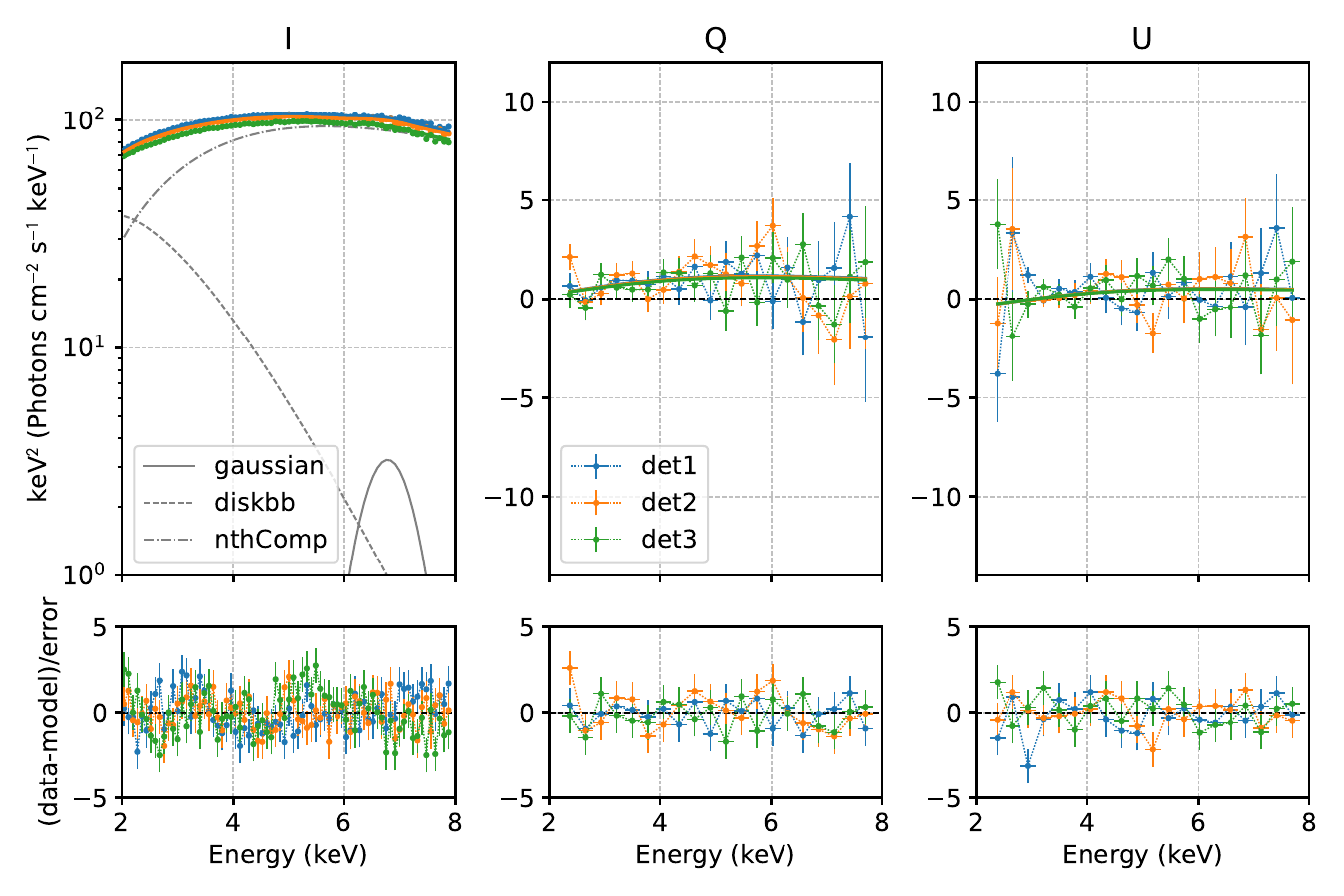}
\caption{Spectral joint fit of the Stokes parameters $I$, $Q$, and $U$ in $EF_E$ representation in the 2--8 keV energy band using the three IXPE detectors and applying the model \texttt{tbabs*(gauss+polconst*diskbb+polconst*nthcomp)}. The best-fit has $\chi^2$/dof = 1329/1337=0.99. Energy rebinning is only for plotting purposes.}
\label{fig:spec_pol}
\end{figure*}

\section{Spectro-polarimetric analysis}\label{sec:spectro-pol}

We analyzed the polarimetric data by using \textsc{xspec} v.12.13.0c \citep{Arnaud96}, firstly freezing the spectral model at the simplest Model A reported in Section~\ref{sec:spectral_model} with the best-fit values reported in Table~\ref{tab:spectrum}. The spectral features due to the reflection, including the broad Fe line, are hardly distinguishable with the \ixpe energy resolution and narrow energy band. When Model A is applied only to the \ixpe $I$ spectra, we get a $\chi^2$/dof = 497/436=1.14. Therefore, we fitted $I$, $Q$, and $U$ spectra simultaneously (applying the same gain corrections to their responses) using the simple model \texttt{tbabs*polconst*(diskbb+nthcomp+gauss)} to compare the spectro-polarimetric results with the one obtained by \texttt{PCUBE}. The polarization in the \ixpe nominal 2--8\,keV energy band is:  PD = 1.0\%$\pm$0.2\% and PA = $8\degr\pm6\degr$ at 90\% of CL with a $\chi^2$/dof = 1333/1337=1.00. When this analysis is applied in the same 3--8 keV energy band of \texttt{PCUBE}, as reported also in Table~\ref{tab:pol_pcube_xspec}, the polarization values are well compatible within 68\% of CL. The same results are obtained when the \ixpe nominal energy band is divided into 1\,keV energy bins (see Table~\ref{tab:pol_pcube_xspec}). This spectro-polarimetric analysis also allows us to determine the polarization values in the 2--3\,keV energy bin, where the \texttt{PCUBE} algorithm overestimates the polarization. The PD and PA values in 2--3\,keV are compatible with the ones at higher energies within 68\% CL, confirming no significant variations of polarization with energy. 

As the iron line emission is not expected to be polarized because of their isotropic emission \citep[see, e.g.,][]{Churazov02}, we fitted the data with a more physical model: \texttt{tbabs*(polconst*(diskbb+nthcomp)+gauss)}, with an unpolarized gaussian line. The result is compatible with the one of the previous model, obtaining in the 2--8\,keV energy band: PD=1.0\% $\pm$ 0.2\% and PA=$8\degr \pm 6\degr$ at 90\% of CL with a $\chi^2$/dof = 1371/1341=1.02. For this model, we also tried to apply a polarization varying linearly with energy by using \texttt{pollin} model instead of \texttt{polconst}; the result is a slope compatible with zero for both PD and PA ($A_{\rm slope}=(0.11\pm0.18)$\%\,keV$^{-1}$, $\psi_{\rm slope}$=(3\degr$\pm$5\degr)\,keV$^{-1}$ at 90\% CL), as expected from the results in the \texttt{PCUBE} analysis, see Table~\ref{tab:pol_pcube_xspec}. In conclusion, model-dependent spectro-polarimetric analysis confirms no evidence of variation of PD or PA in the \ixpe nominal energy band.  

To complete our spectro-polarimetric analysis, we applied a different model with \texttt{polconst} associated at each spectral component, considering an unpolarized Gaussian line: \texttt{tbabs*(polconst*diskbb+polconst*nthcomp+gauss)}.
This model can allow us to disentangle polarization of the disk emission from the one due to Comptonization in the SL/BL. The results for this fit are reported in Table~\ref{tab:specpol} and Figure~\ref{fig:spec_pol}. For the disk component, we obtained a PD of $1.5\%\pm1.0\%$ and a PA of $-$40\degr$\pm$21\degr\ at  68\% CL, corresponding to an upper limit for the PD of $<3.2\%$ at 90\% CL. For the Comptonization component, we obtained a PD of $1.3\%\pm0.4\%$  and a PA of 14\degr$\pm$8\degr\  at 90\% CL.

\begin{deluxetable}{llc} 
\tablecaption{Best-fit parameters for the spectro-polarimetric analysis with the model \texttt{polconst*diskbb+polconst*nthcomp+gauss}.}
\label{tab:specpol}
\tablehead{ 
\colhead{Component} & & \colhead{Value} 
 }
\startdata
\texttt{diskbb} & PD (\%) & $<3.2$  \\
    & PA ($\deg$) & --- \\ \hline
\texttt{nthcomp} & PD (\%) & $1.3 \pm 0.4$ \\ 
    & PA ($\deg$)  & $14\pm8$ \\
    \hline
    & $\chi^2$/dof & 1329/1337 = 0.99 \\ 
\enddata
\tablecomments{Errors are at 90\% CL. }
\end{deluxetable}

As a further step, we tried to apply the spectro-polarimetric analysis to Model C, including modeling for the reflection component, aiming to understand its contribution to the total polarization better. Firstly, we associate a different constant polarization to each spectral component, but due to the limited spectral capabilities of \ixpe, a clear polarimetric disentanglement of the three components in the $Q$ and $U$ spectra is not possible. As a result, just upper limits at 90\% CL of $<1.9\%$ for the disk component, $<8.2\%$ for the Comptonized component, and $<66\%$ for the reflection one can be obtained. Then, we fixed the PD and PA for the disk and Comptonization components to the values of Table~\ref{tab:specpol}, obtaining a PD of the reflection $<6.4\%$, still unconstrained at 90\% CL. As a latter scenario, we tried to fix the PD of the diskbb at 1.1\%, corresponding to the classical results of \citet{Chandrasekhar1960} for inclination of 44\degr, with PA at $-40\degr$, and we assumed an unpolarized Comptonization component. 
In this case, we obtained a PD=$14\%\pm5\%$ and PA=$15\degr\pm7\degr$ (errors at 90\% CL) for the reflection.

\section{Discussion}

\ixpe observed \sco mainly in the SA state with short periods in the FB, as shown by the \nustar CCD of Figure~\ref{fig:CCD}, and also by the presence of QPOs in \nicer data (see Figure~\ref{fig:pds}). As shown in Figure~\ref{fig:contour_flaring}, no evidence for variation of polarization between these two states is observed. The average polarization, representative of the SA state, is measured to be PD$=1.1\% \pm0.2\%$ and PA$=10\degr\pm7\degr$ in the 3--8\,keV energy band by the model-independent \texttt{PCUBE} analysis (significance $\sim$7$\sigma$ CL). The \textsc{xspec} spectro-polarimetric analysis in the whole \ixpe 2--8\,keV nominal energy band applied to the simple Model A of Table~\ref{tab:spectrum}, assuming an unpolarized Gaussian, results in an average PD of 1.0\%$\pm$0.2\%  and a PA of $8\degr \pm 6\degr$ (errors at 90\% CL). 

An energy resolved analysis, applying a 1\,keV binning has been performed, and no significant evidence for variation with energy of the PD or PA is observed either in the \texttt{PCUBE} analysis or the \textsc{xspec} one. Also, an attempt to fit data with the \texttt{pollin} \textsc{xspec} model confirms PD and PA to be constant in the whole \ixpe nominal 2--8\,keV energy band. This result is different, e.g., from the previous one obtained by \ixpe for the ultra-compact 4U~1820$-$303 \citep{DiMarco23}. The observed constant polarization, even if smaller than expectations, can be compatible with a corona geometry, as in the case C of \citet{Poutanen_2023} proposed for a BH in the hard-state, where the hot medium is situated within the truncated cold accretion disk with the seed unpolarized photons coming from the outer cold truncated disk having incident angles limited by the aspect ratio of the flow. 
Although this geometry is not compatible with the present observation where \sco is in a soft state and the disk approaches the NS surface, the most important aspect of the model is the angular distribution of the incident seed photons for scattering. 
Thus, this model can give some indication of a geometry for which no variation of PD/PA with energy is expected. 
Alternative scenarios are discussed in \citet{Schnittman2010}, where such behavior is expected for a sandwich corona in AGNs, and in \citet{gnarini2022} for a sandwich/wedge and/or a shell corona.

When a different polarization is associated to each spectral component of Model A, we obtain an upper limit of $<$3.2\%  at 90\% CL for the PD of the disk component, compatible with expectations for an electron-scattering dominated optically thick accretion disk \citep{Chandrasekhar1960, Sobolev63} that is $\sim$1.1\% for an inclination of $\sim$44\degr\ as measured for \sco \citep{Fomalont01}. For the Comptonized component, we measured a PD of $1.3\%\pm0.4\%$, compatible with the result obtained for a Thomson optical depth $\tau_{\rm T}$$\sim$7 and $kT_{\rm e}$$\sim$3~keV \citep[see Figure~5 in][]{st85}. 

\sco spectral analysis of the multi-observatory observations also allow for the study of the reflection component. In \citet{349+2}, a study of the reflection in the Sco-like source \mbox{GX 349+2}, showed a strong agreement between different states along the Z-track, except for the extreme flaring branch, this agrees with the good reduced $\chi^2$ we obtain mixing the SA state and the short flaring states in the present observations. For \mbox{GX 349+2} they reported an average inner radius of about 33\,km that is not varying in the different states, except the extreme flaring; this $R_{\rm in}$ value is larger with respect to the present estimate obtained for \sco, which is $\sim$18\,km for a canonical $1.4M_\odot$ NS. This means that in the present \sco observation, the average inner radius is closer to the NS surface, which can be considered an upper limit on the NS radius. An alternative estimate of the inner disk radius can also be obtained by the \texttt{diskbb} component used to describe the continuum in the different models of Table~\ref{tab:spectrum}; in the Model C, when from the \texttt{diskbb} result for non-zero-torque boundary condition we obtain a slightly larger $R_{\rm in}$, at level of $13R_{\rm g}$, but still confirming a disk approaching the NS surface.

The \textsc{xspec} spectro-polarimetric analysis has also been applied to Model C, which includes modeling of the reflection component. To achieve a constrained polarization estimate for the reflection component, we need to fix the PD for the disk at 1.1\% (corresponding to the classical results of \citealt{Chandrasekhar1960} for $i=44\degr$) with PA=$-40\degr$, and to assume an unpolarized Comptonization component, that is almost the case reported in Figure~5 of \citet{st85} for the optical thickness $\sim$7 (see Table~\ref{tab:spectrum}). In this scenario, we obtain a PD=$14\%\pm5\%$ and PA=$15\degr\pm7\degr$ (errors at 90\% CL), compatible with the predicted polarization values for a Compton-reflected spectrum from cold matter, as reported by \citet{Matt93} and \citet{Poutanen96}.

\sco is a bright radio source, where radio-jets were clearly identified \citep{Bradshaw99,Andrew68,Fomalont01}. These radio-jets are connected to the accretion flow, but the properties of the jet and their correlation with the accretion flow are not so well understood \citep{Motta19}. Since the X-ray PA is strongly dependent on the geometry of the accretion flow, the correlation between the radio-jet position angle and the PA is of great interest. For example, a previous study of the polarization in \sco, reported in \citet{Long2022}, was interpreted as an indication for a \sco corona residing in a vertically-extended BL/SL located between the inner disk and the NS. In fact, for this scenario, the X-ray PA is expected to be perpendicular to the disk, and aligned with the system axis as the radio-jet is assumed to be. Similarly, the \ixpe results obtained for Cyg X-2 \citep{Farinelli23} and in the stellar-mass BH Cyg X-1 \citep{Krawczynski2022} show an indication for a PA aligned with the radio-jet. In the measurements reported here, both the model-independent and the spectro-polarimetric analysis show a PA at $8\degr \pm 6\degr$, that is not aligned with the \sco radio-jet which is at $54\degr \pm 3\degr$. The resulting rotation of the PA with respect to the radio-jet position angle is $46\degr \pm 9\degr$, different from the previous indication of alignment between PA and radio-jet. However, none of these X-ray polarization measurements were performed simultaneously with the radio ones. The orientation of the jet in a few sources has been seen to change on short timescales (hours), as in the case of the BH system V404 Cygni \citep{Miller-Jones2019}, likely due to the Lense-Thirring precession of the accretion disk \citep{Stella98}. Although not yet observed, this can also be expected for NS, and the observed misalignment between \ixpe measured PA and radio-jet \citep{Fomalont2001b} could be consistent with this effect (e.g. the V404 Cygni rapid variation of the radio-jet position angle ranged from $-30\fdg6$ up to $+5\fdg6$, that is a 36$\degr$ rotation).

This new result by \ixpe is a strong improvement in terms of significance with respect to the two previous attempts to measure the X-ray polarization of \sco. The first one was performed by OSO-8 \citep{Long1979}, and the second one was recently performed by PolarLight \citep{Long2022}. Neither of these attempts were performed with an instrument capable of performing spectral analysis; thus, it was not possible to identify the status of \sco. Moreover, only upper limits or marginal detections were achieved. OSO-8 PD is compatible within 1$\sigma$ with the present ones of Table~\ref{tab:pol_pcube_xspec} at both 2.6 and  5.2\,keV. The PA shows marginal variations with respect to this \ixpe measurement at levels of $33\degr\pm14\degr$ and $45\degr \pm 8\degr$ at 2.6 and 5.2 keV, respectively. PolarLight divided its data in two energy bins and two source flux states, low and high, on the basis of the measured count rate. For the low-flux state, PolarLight was capable of obtaining only upper limits in both energy bins that are compatible with the \ixpe result. For the high-flux state: the 3--4 keV PD is still compatible, while the PA differs by $55\degr \pm 19 \degr$; the 4--8 keV PD differs by 3.0\%$\pm$0.8\% and the PA differs by $41\degr \pm 7\degr$. When PolarLight data are analyzed without any flux selection, the upper limit in 3--4 keV is compatible with the \ixpe result, while the 4--8 keV values differ by 1.3\%$\pm$0.6\% for the PD and $44\degr \pm 9\degr$ for the PA. 

This comparison with previous observations, although without a clear state identification, in light of previous \ixpe results for other Z-sources \citep{Farinelli23,Cocchi2023, Fabiani23,Rankin23} can allow for some considerations. In \citet{Cocchi2023}, \citet{Fabiani23}, and \citet{Rankin23}, variations of the polarimetric properties along the Z track are reported. The sources observed in the HB \citep{Farinelli23,Cocchi2023,Fabiani23} showed a higher PD, in line with the result of PolarLight for the high flux \citep{Long2022}. Moreover, for Cyg X-2 \citep{Farinelli23} the HB PA results to be aligned with the measured radio-jet direction. On the other hand, this \sco observation was dominated by the SA, where the flux is smaller, and this may correspond to the low-flux state in the PolarLight result, which is compatible, although with a much smaller significance, with the \ixpe result. These considerations give an indication of a polarization variation along the Z branches for \sco with a NB/HB state having higher polarization, at level of the PolarLight result, and PA variations up to $\sim$50\degr\ in line with other \ixpe results \citep{Cocchi2023,Rankin23}. \ixpe's different measured PA with respect to the previous observations can be explained with a variation of the corona geometry in the different states of the source. 

Because of these uncertainties on how the polarization changes along the Z branches, it may be crucial to have new \sco observations, possibly coordinated with radio ones, to correlate the jet emission direction with the PA. This kind of study could be of great impact, allowing us to verify the impact of the Lense-Thirring precession and/or geometrical variations of the radio emissions with the corona geometry.

\section{Summary}

In this paper, we report the highly significant ($\sim$7$\sigma$ CL) detection of polarization from \sco obtained by \ixpe. During the \ixpe observation, the source was observed mainly in the Soft Apex state. The spectral analysis has been performed using simultaneous \nicer, \nustar, \hxmt, and \ixpe observations. The spectral model is described by \texttt{diskbb} to describe the direct disk or SL emission, \texttt{nthcomp} to describe the Comptonization component having blackbody seed photons emitted by the neutron star BL/SL, and a broad line at $\sim$6.7\,keV due to reflection (also more complex reflection model can be included). The reflection analysis by using the \texttt{relxillNS} model improves the reduced $\chi^2$  and allows setting an upper limit on the inner disk radius of $\sim$2\,$R_{\rm ISCO}$, compatible with a scenario with the disk close the NS surface. A PD at level of 1\% is measured, and a PA misalignment with respect to the direction of the radio-jet, that is $54\degr$ \citep[north to east on the sky plane,][]{Fomalont01,Fomalont2001b}, is observed. A discrepancy between \ixpe results and the previous attempts by OSO-8 and PolarLight is also reported. 

Polarization studies in other Z-sources, performed by \ixpe \citep{Cocchi2023,Fabiani23,Rankin23}, showed polarization variations along the different states of the sources with PA changing up to $\sim$50$\degr$. This can explain these PA discrepancies in terms of changes of the Z branch in which \sco was observed. In conclusion, these different PA can indicate a corona geometry that changes along the Z-track. Alternatively, the \ixpe PA misalignment with previous radio-jet direction measurements can be explained with relativistic precession that may have altered the radio-jet position. 


\section*{Acknowledgments}

The Imaging X-ray Polarimetry Explorer (IXPE) is a joint US and Italian mission.  The US contribution is supported by the National Aeronautics and Space Administration (NASA) and led and managed by its Marshall Space Flight Center (MSFC), with industry partner Ball Aerospace (contract NNM15AA18C).  The Italian contribution is supported by the Italian Space Agency (Agenzia Spaziale Italiana, ASI) through contract ASI-OHBI-2022-13-I.0, agreements ASI-INAF-2022-19-HH.0 and ASI-INFN-2017.13-H0, and its Space Science Data Center (SSDC) with agreements ASI-INAF-2022-14-HH.0 and ASI-INFN 2021-43-HH.0, and by the Istituto Nazionale di Astrofisica (INAF) and the Istituto Nazionale di Fisica Nucleare (INFN) in Italy.  This research used data products provided by the IXPE Team (MSFC, SSDC, INAF, and INFN) and distributed with additional software tools by the High-Energy Astrophysics Science Archive Research Center (HEASARC), at NASA Goddard Space Flight Center (GSFC).

The authors acknowledge the NICER, \nustar, and \hxmt teams for the prompt scheduling of the simultaneous observations. In particular, the authors acknowledge Karl Forster, Hannah Penn Earnshaw, Hiromasa Miyasaka, Murray Brightman, and Fiona A. Harrison.

We acknowledge support from the Academy of Finland grants 333112, 355672, and  349144 (JP, AV, SST) and the German Academic Exchange Service (DAAD) travel grant 57525212 (VD). 
APap, FA, and GI are supported by INAF (Research Grant ''Uncovering the optical beat of the fastest magnetised neutron stars (FANS)'') and the Italian Ministry of University and Research (MUR) (PRIN 2020, Grant 2020BRP57Z, ''Gravitational and Electromagnetic-wave Sources in the Universe with current and next-generation detectors (GEMS)'').
FX is supported by the National Natural Science Foundation of China (Grant No. 12373041). 
KL is supported by the National Natural Science Foundation of China (Grant No. 12133003).
IL was supported by the NASA Postdoctoral Program at the Marshall Space Flight Center, administered by Oak Ridge Associated Universities under contract with NASA.
POP acknowledges financial support from the French High Energy Program (PNHE/CNRS) and from the French Space Agency (CNES).


%

\vspace{5mm}
\facilities{\ixpe, \nicer, \nustar, \hxmt}


\software{\textsc{ixpeobssim} \citep{Baldini22}, \textsc{xspec} \citep{Arnaud96}, \textsc{HEASoft} \citep{1995ASPC...77..367B}, \textsc{stingray} \citep{huppenkothen2019}, \textsc{HXMTDAS} \citep{Zhang20}}

\appendix


\section{Data handling for X-ray observatories}\label{sec:dataX}
In the following sub-sections, we report how data were extracted, and the software tools we used for the analysis of the data from X-ray observatories.

\subsection{\ixpe}

The Imaging X-ray Polarimetry Explorer (\ixpe) is a joint NASA and Italian Space Agency mission launched on 2021 December 9 and entirely dedicated to measuring linear X-ray polarization. The performance and an extensive description of the observatory are reported in \citet{Weisskopf2022}. \ixpe comprises three identical grazing incidence telescopes, providing imaging and spectral polarimetry over the 2--8\,keV energy band with a time resolution better than 15~$\mu$s. Each telescope couples an X-ray module of mirror assembly (MMA) and a detector unit (DU), hosting a gas-pixel detector (GPD), which is a photoelectric polarimeter sensitive to linear X-ray polarization \citep{Costa2001, Soffitta21}. 

Thanks to its imaging capability, we selected the source photons from the \ixpe telescope images inside a circular region of radius 100\arcsec\ centered on the source position. No background regions were selected on the images, following the prescription reported in \citet{Di_Marco2023} for bright sources with a high count rate ($\gtrsim2$~cnt s$^{-1}$),  which is the case of \sco. 

The \ixpe data were processed using 
\texttt{xpbin} tool in  \textsc{ixpeobssim} \citep{Baldini22} to obtain the PHA Stokes spectra for the spectro-polarimetric analysis with \textsc{xspec}. All obtained spectra were grouped to have at least 50 counts per bin.

\subsection{NICER}

The Neutron Star Interior Composition Explorer -- NICER -- onboard the International Space Station (ISS) and launched in June 2017, is a soft X-ray instrument consisting of 56 co-aligned concentrator X-ray optics, each one having a single silicon drift detector in the focus \citep{Gendreau16}. NICER does not offer imaging capabilities but has a large collecting area providing unmatched time resolution in the soft X-ray bandpass, with a sensitive energy interval 0.2–12\,keV. Contemporaneous observations of \sco during \ixpe observation were performed in the framework of the NICER GO Cycle 5 (proposal 6189).

The NICER data were processed with the NICER Data Analysis Software v010a released on 2022 December 16 provided under \textsc{HEASoft} v6.31.1 with the CALDB version released on 2022 October 30. The background spectra have been estimated by applying the new SCORPEON model. All obtained spectra were grouped to have at least 50 counts per bin.

\subsection{NuSTAR}

\nustar, the Nuclear Spectroscopic Telescope Array, observatory \citep{Harrison2013}, consists of two identical X-ray telescopes, referred to as FPMA and FPMB. They provide broadband X-ray imaging, spectroscopy, and timing in the energy range 3--79\,keV with angular resolution 18\arcsec\ (FWHM) and spectral resolution of 400\,eV (FWHM) at 10\,keV. 
The \nustar observation was performed in the framework of the \nustar GO Cycle 9 (proposal 9212).

The \nustar data were processed with the standard Data Analysis Software (\texttt{nustardas} 16Feb22 v2.1.2) provided under \textsc{HEASoft} v6.31.1 with the CALDB version released on 2023 April 4 by using the \texttt{statusexpr="STATUS==b0000xxx00xxxx000"} keyword in the \textsc{nupipeline}, as suggested for bright sources\footnote{\href{https://heasarc.gsfc.nasa.gov/docs/nustar/analysis/}{https://heasarc.gsfc.nasa.gov/docs/nustar/analysis/}}. Both source and background regions are obtained by applying a circular 150\arcsec\ radius selection; the spectra are extracted in these regions centered on the location of \sco for the source, and, for the background, in an off-center sourceless region in the field of view. All the obtained spectra were grouped to have at least 50 counts per bin.

\subsection{\hxmt}
The Hard X-ray Modulation Telescope (\hxmt) is the first Chinese X-ray satellite launched in June 2017 \citep{Zhang14, Zhang20}. The satellite comprises three different instruments: (i) the low-energy (LE) instrument, working in 1--15\,keV and consisting of 96 swept charge devices, with an effective area of 384\,cm$^2$; (ii) the medium-energy (ME)  instrument, working in the 5--30\,keV range and consisting of 1728 Si-Pin units, with an effective area of 952\,cm$^2$; (iii) the high-energy (HE) instrument, working in 20--250\,keV and consisting of 18 NaI(TI)/CsI(Na) scintillation detectors, with an effective area of 5100\,cm$^2$ \citep{Li20,Zhang20}.

The data reduction of \hxmt was performed utilizing \hxmt Data Analysis Software \texttt{HXMTDAS} v2.04 with default filter criteria, including elevation angle (ELV) $>$ 10$^{\circ}$, a geometric cutoff rigidity (COR) $>$ 8\,GeV, offset for the point position $\leq$ 0.04$^{\circ}$, and time beyond 300\,s to the South Atlantic Anomaly (SAA).



%


\bibliography{biblio}{}

\begin{thebibliography}{}
\expandafter\ifx\csname natexlab\endcsname\relax\def\natexlab#1{#1}\fi
\providecommand{\url}[1]{\href{#1}{#1}}
\providecommand{\dodoi}[1]{doi:~\href{http://doi.org/#1}{\nolinkurl{#1}}}
\providecommand{\doeprint}[1]{\href{http://ascl.net/#1}{\nolinkurl{http://ascl.net/#1}}}
\providecommand{\doarXiv}[1]{\href{https://arxiv.org/abs/#1}{\nolinkurl{https://arxiv.org/abs/#1}}}

\bibitem[{{Andrew} \& {Purton}(1968)}]{Andrew68}
{Andrew}, B.~H., \& {Purton}, C.~R. 1968, \nat, 218, 855,
  \dodoi{10.1038/218855a0}

\bibitem[{{Arnason} {et~al.}(2021){Arnason}, {Papei}, {Barmby}, {Bahramian}, \&
  {Gorski}}]{Arnason21}
{Arnason}, R.~M., {Papei}, H., {Barmby}, P., {Bahramian}, A., \& {Gorski},
  M.~D. 2021, \mnras, 502, 5455, \dodoi{10.1093/mnras/stab345}

\bibitem[{{Arnaud}(1996)}]{Arnaud96}
{Arnaud}, K.~A. 1996, in ASP Conf. Ser., Vol. 101, Astronomical Data Analysis
  Software and Systems V, ed. G.~H. {Jacoby} \& J.~{Barnes} (San Francisco:
  Astron. Soc. Pac.), 17--20

\bibitem[{{Bahramian} \& {Degenaar}(2023)}]{Bahramian2022}
{Bahramian}, A., \& {Degenaar}, N. 2023, in Handbook of X-ray and Gamma-ray
  Astrophysics, ed. C.~{Bambi} \& A.~{Santangelo} (Singapore: Springer Nature),
  120, \dodoi{10.1007/978-981-16-4544-0_94-1}

\bibitem[{{Baldini} {et~al.}(2022){Baldini}, {Bucciantini}, {Lalla}, {Ehlert},
  {Manfreda}, {Negro}, {Omodei}, {Pesce-Rollins}, {Sgr{\`o}}, \&
  {Silvestri}}]{Baldini22}
{Baldini}, L., {Bucciantini}, N., {Lalla}, N.~D., {et~al.} 2022, SoftwareX, 19,
  101194, \dodoi{10.1016/j.softx.2022.101194}

\bibitem[{Barret \& Vaughan(2012)}]{barret2012}
Barret, D., \& Vaughan, S. 2012, ApJ, 746, 131,
  \dodoi{10.1088/0004-637X/746/2/131}

\bibitem[{Bartlett(1948)}]{bartlett1948}
Bartlett, M.~S. 1948, Nature, 161, 686, \dodoi{10.1038/161686a0}

\bibitem[{{Blackburn}(1995)}]{1995ASPC...77..367B}
{Blackburn}, J.~K. 1995, in ASP Conf. Ser., Vol.~77, Astronomical Data Analysis
  Software and Systems IV, ed. R.~A. {Shaw}, H.~E. {Payne}, \& J.~J.~E. {Hayes}
  (San Francisco: Astron. Soc. Pac.), 367

\bibitem[{{Bradshaw} {et~al.}(1999){Bradshaw}, {Fomalont}, \&
  {Geldzahler}}]{Bradshaw99}
{Bradshaw}, C.~F., {Fomalont}, E.~B., \& {Geldzahler}, B.~J. 1999, \apjl, 512,
  L121, \dodoi{10.1086/311889}

\bibitem[{{Cackett} {et~al.}(2012){Cackett}, {Miller}, {Reis}, {Fabian}, \&
  {Barret}}]{Cackett12}
{Cackett}, E.~M., {Miller}, J.~M., {Reis}, R.~C., {Fabian}, A.~C., \& {Barret},
  D. 2012, \apj, 755, 27, \dodoi{10.1088/0004-637X/755/1/27}

\bibitem[{{Cackett} {et~al.}(2008){Cackett}, {Miller}, {Bhattacharyya},
  {Grindlay}, {Homan}, {van der Klis}, {Miller}, {Strohmayer}, \&
  {Wijnands}}]{Cackett08}
{Cackett}, E.~M., {Miller}, J.~M., {Bhattacharyya}, S., {et~al.} 2008, \apj,
  674, 415, \dodoi{10.1086/524936}

\bibitem[{{Cackett} {et~al.}(2010){Cackett}, {Miller}, {Ballantyne}, {Barret},
  {Bhattacharyya}, {Boutelier}, {Miller}, {Strohmayer}, \&
  {Wijnands}}]{Cackett10}
{Cackett}, E.~M., {Miller}, J.~M., {Ballantyne}, D.~R., {et~al.} 2010, \apj,
  720, 205, \dodoi{10.1088/0004-637X/720/1/205}

\bibitem[{{Capitanio} {et~al.}(2023){Capitanio}, {Fabiani}, {Gnarini},
  {Ursini}, {Ferrigno}, {Matt}, {Poutanen}, {Cocchi}, {Mikusincova},
  {Farinelli}, {Bianchi}, {Kajava}, {Muleri}, {Sanchez-Fernandez}, {Soffitta},
  {Wu}, {Agudo}, {Antonelli}, {Bachetti}, {Baldini}, {Baumgartner},
  {Bellazzini}, {Bongiorno}, {Bonino}, {Brez}, {Bucciantini}, {Castellano},
  {Cavazzuti}, {Ciprini}, {Costa}, {De Rosa}, {Del Monte}, {Di Gesu}, {Di
  Lalla}, {Di Marco}, {Donnarumma}, {Doroshenko}, {Dov{\v{c}}iak}, {Ehlert},
  {Enoto}, {Evangelista}, {Ferrazzoli}, {Garcia}, {Gunji}, {Hayashida}, {Heyl},
  {Iwakiri}, {Jorstad}, {Karas}, {Kitaguchi}, {Kolodziejczak}, {Krawczynski},
  {La Monaca}, {Latronico}, {Liodakis}, {Maldera}, {Manfreda}, {Marin},
  {Marinucci}, {Marscher}, {Marshall}, {Mitsuishi}, {Mizuno}, {Ng}, {O'Dell},
  {Omodei}, {Oppedisano}, {Papitto}, {Pavlov}, {Peirson}, {Perri},
  {Pesce-Rollins}, {Petrucci}, {Pilia}, {Possenti}, {Puccetti}, {Ramsey},
  {Rankin}, {Ratheesh}, {Romani}, {Sgr{\`o}}, {Slane}, {Spandre}, {Tamagawa},
  {Tavecchio}, {Taverna}, {Tawara}, {Tennant}, {Thomas}, {Tombesi}, {Trois},
  {Tsygankov}, {Turolla}, {Vink}, {Weisskopf}, {Xie}, \& {Zane}}]{Capitanio23}
{Capitanio}, F., {Fabiani}, S., {Gnarini}, A., {et~al.} 2023, \apj, 943, 129,
  \dodoi{10.3847/1538-4357/acae88}

\bibitem[{{Casella} {et~al.}(2006){Casella}, {Belloni}, \&
  {Stella}}]{Casella06}
{Casella}, P., {Belloni}, T., \& {Stella}, L. 2006, \aap, 446, 579,
  \dodoi{10.1051/0004-6361:20052912}

\bibitem[{{Chandrasekhar}(1960)}]{Chandrasekhar1960}
{Chandrasekhar}, S. 1960, {Radiative Transfer} (New York: Dover)

\bibitem[{Chatterjee {et~al.}(2023)Chatterjee, Agrawal, Jayasurya, \&
  Katoch}]{Chatterjee2023}
Chatterjee, R., Agrawal, V.~K., Jayasurya, K.~M., \& Katoch, T. 2023, \mnras,
  521, L74, \dodoi{10.1093/mnrasl/slad026}

\bibitem[{{Churazov} {et~al.}(2002){Churazov}, {Sunyaev}, \&
  {Sazonov}}]{Churazov02}
{Churazov}, E., {Sunyaev}, R., \& {Sazonov}, S. 2002, \mnras, 330, 817,
  \dodoi{10.1046/j.1365-8711.2002.05113.x}

\bibitem[{{Church} {et~al.}(2012){Church}, {Gibiec},
  {Ba{\l}uci{\'n}ska-Church}, \& {Jackson}}]{Church12}
{Church}, M.~J., {Gibiec}, A., {Ba{\l}uci{\'n}ska-Church}, M., \& {Jackson},
  N.~K. 2012, \aap, 546, A35, \dodoi{10.1051/0004-6361/201218987}

\bibitem[{{Cocchi} {et~al.}(2023){Cocchi}, {Gnarini}, {Fabiani}, {Ursini},
  {Poutanen}, {Capitanio}, {Bobrikova}, {Farinelli}, {Paizis}, {Sidoli},
  {Veledina}, {Bianchi}, {Di Marco}, {Ingram}, {Kajava}, {La Monaca}, {Matt},
  {Malacaria}, {Miku{\v{s}}incov{\'a}}, {Rankin}, {Zane}, {Agudo}, {Antonelli},
  {Bachetti}, {Baldini}, {Baumgartner}, {Bellazzini}, {Bongiorno}, {Bonino},
  {Brez}, {Bucciantini}, {Castellano}, {Cavazzuti}, {Chen}, {Ciprini}, {Costa},
  {De Rosa}, {Del Monte}, {Di Gesu}, {Di Lalla}, {Donnarumma}, {Doroshenko},
  {Dov{\v{c}}iak}, {Ehlert}, {Enoto}, {Evangelista}, {Ferrazzoli}, {Garcia},
  {Gunji}, {Hayashida}, {Heyl}, {Iwakiri}, {Jorstad}, {Kaaret}, {Karas},
  {Kislat}, {Kitaguchi}, {Kolodziejczak}, {Krawczynski}, {Latronico},
  {Liodakis}, {Maldera}, {Manfreda}, {Marin}, {Marinucci}, {Marscher},
  {Marshall}, {Massaro}, {Mitsuishi}, {Mizuno}, {Muleri}, {Negro}, {Ng},
  {O'Dell}, {Omodei}, {Oppedisano}, {Papitto}, {Pavlov}, {Peirson}, {Perri},
  {Pesce-Rollins}, {Petrucci}, {Pilia}, {Possenti}, {Puccetti}, {Ramsey},
  {Ratheesh}, {Roberts}, {Romani}, {Sgr{\`o}}, {Slane}, {Soffitta}, {Spandre},
  {Swartz}, {Tamagawa}, {Tavecchio}, {Taverna}, {Tawara}, {Tennant}, {Thomas},
  {Tombesi}, {Trois}, {Tsygankov}, {Turolla}, {Vink}, {Weisskopf}, {Wu}, \&
  {Xie}}]{Cocchi2023}
{Cocchi}, M., {Gnarini}, A., {Fabiani}, S., {et~al.} 2023, \aap, 674, L10,
  \dodoi{10.1051/0004-6361/202346275}

\bibitem[{Costa {et~al.}(2001)Costa, Soffitta, Bellazzini, Brez, Lumb, \&
  Spandre}]{Costa2001}
Costa, E., Soffitta, P., Bellazzini, R., {et~al.} 2001, Nature, 411, 662,
  \dodoi{10.1038/35079508}

\bibitem[{{Coughenour} {et~al.}(2018){Coughenour}, {Cackett}, {Miller}, \&
  {Ludlam}}]{349+2}
{Coughenour}, B.~M., {Cackett}, E.~M., {Miller}, J.~M., \& {Ludlam}, R.~M.
  2018, \apj, 867, 64, \dodoi{10.3847/1538-4357/aae098}

\bibitem[{{Di Marco} {et~al.}(2022{\natexlab{a}}){Di Marco}, Costa, Muleri,
  Soffitta, Fabiani, {La Monaca}, Rankin, Xie, Bachetti, Baldini, Baumgartner,
  Bellazzini, Brez, Castellano, {Del Monte}, {Di Lalla}, Ferrazzoli, Latronico,
  Maldera, Manfreda, O’Dell, Perri, Pesce-Rollins, Puccetti, Ramsey,
  Ratheesh, Sgrò, Spandre, Tennant, Tobia, Trois, \& Weisskopf}]{DiMarco_2022}
{Di Marco}, A., Costa, E., Muleri, F., {et~al.} 2022{\natexlab{a}}, \aj, 163,
  170, \dodoi{10.3847/1538-3881/ac51c9}

\bibitem[{{Di Marco} {et~al.}(2022{\natexlab{b}}){Di Marco}, {Muleri},
  {Fabiani}, {La Monaca}, {Rankin}, {Soffitta}, {Baldini}, {Costa}, {Del
  Monte}, {Ferrazzoli}, {Lefevre}, {Maiolo}, {Maita}, {Manfreda}, {Morbidini},
  {O'Dell}, {Ramsey}, {Ratheesh}, {Sgro'}, {Trois}, {Tennant}, \&
  {Weisskopf}}]{DiMarco_2022b}
{Di Marco}, A., {Muleri}, F., {Fabiani}, S., {et~al.} 2022{\natexlab{b}}, in
  \procspie, Vol. 12181, Space Telescopes and Instrumentation 2022: Ultraviolet
  to Gamma Ray, ed. J.-W.~A. {den Herder}, S.~{Nikzad}, \& K.~{Nakazawa},
  121811C, \dodoi{10.1117/12.2629413}

\bibitem[{{Di Marco} {et~al.}(2023{\natexlab{a}}){Di Marco}, {La Monaca},
  {Poutanen}, {Russell}, {Anitra}, {Farinelli}, {Mastroserio}, {Muleri}, {Xie},
  {Bachetti}, {Burderi}, {Carotenuto}, {Del Santo}, {Di Salvo},
  {Dov{\v{c}}iak}, {Gnarini}, {Iaria}, {Kajava}, {Liu}, {Middei}, {O'Dell},
  {Pilia}, {Rankin}, {Sanna}, {Eijnden}, {Weisskopf}, {Bobrikova}, {Capitanio},
  {Costa}, {Kaaret}, {Marino}, {Soffitta}, {Ursini}, {Ambrosino}, {Cocchi},
  {Fabiani}, {Marshall}, {Matt}, {Motta}, {Papitto}, {Stella}, {Tarana},
  {Zane}, {Agudo}, {Antonelli}, {Baldini}, {Baumgartner}, {Bellazzini},
  {Bianchi}, {Bongiorno}, {Bonino}, {Brez}, {Bucciantini}, {Castellano},
  {Cavazzuti}, {Chen}, {Ciprini}, {De Rosa}, {Del Monte}, {Di Gesu}, {Di
  Lalla}, {Donnarumma}, {Doroshenko}, {Ehlert}, {Enoto}, {Evangelista},
  {Ferrazzoli}, {Garcia}, {Gunji}, {Hayashida}, {Heyl}, {Iwakiri}, {Jorstad},
  {Karas}, {Kislat}, {Kitaguchi}, {Kolodziejczak}, {Krawczynski}, {Latronico},
  {Liodakis}, {Maldera}, {Manfreda}, {Marin}, {Marinucci}, {Marscher},
  {Massaro}, {Mitsuishi}, {Mizuno}, {Negro}, {Ng}, {Omodei}, {Oppedisano},
  {Pavlov}, {Peirson}, {Perri}, {Pesce-Rollins}, {Petrucci}, {Possenti},
  {Puccetti}, {Ramsey}, {Ratheesh}, {Roberts}, {Romani}, {Sgr{\`o}}, {Slane},
  {Spandre}, {Swartz}, {Tamagawa}, {Tavecchio}, {Taverna}, {Tawara}, {Tennant},
  {Thomas}, {Tombesi}, {Trois}, {Tsygankov}, {Turolla}, {Vink}, {Wu}, \& {IXPE
  Collaboration}}]{DiMarco23}
{Di Marco}, A., {La Monaca}, F., {Poutanen}, J., {et~al.} 2023{\natexlab{a}},
  \apjl, 953, L22, \dodoi{10.3847/2041-8213/acec6e}

\bibitem[{{Di Marco} {et~al.}(2023{\natexlab{b}}){Di Marco}, {Soffitta},
  {Costa}, {Ferrazzoli}, {La Monaca}, {Rankin}, {Ratheesh}, {Xie}, {Baldini},
  {Del Monte}, {Ehlert}, {Fabiani}, {Kim}, {Muleri}, {O'Dell}, {Ramsey},
  {Rubini}, {Sgr{\`o}}, {Silvestri}, {Tennant}, \& {Weisskopf}}]{Di_Marco2023}
{Di Marco}, A., {Soffitta}, P., {Costa}, E., {et~al.} 2023{\natexlab{b}}, \aj,
  165, 143, \dodoi{10.3847/1538-3881/acba0f}

\bibitem[{{Dickey} \& {Lockman}(1990)}]{Dickey90}
{Dickey}, J.~M., \& {Lockman}, F.~J. 1990, \araa, 28, 215,
  \dodoi{10.1146/annurev.aa.28.090190.001243}

\bibitem[{{Ding} {et~al.}(2023){Ding}, {Qu}, {Song}, {Huang}, {Zhang}, {Bu},
  {Ge}, {Li}, {Tao}, {Ma}, {Chen}, {Zhang}, {Yan}, {Tuo}, {Fu}, {Xiao}, {Yang},
  \& {Liu}}]{Ding23}
{Ding}, G.~Q., {Qu}, J.~L., {Song}, L.~M., {et~al.} 2023, \apj, 950, 69,
  \dodoi{10.3847/1538-4357/accf91}

\bibitem[{{Fabian} {et~al.}(1989){Fabian}, {Rees}, {Stella}, \&
  {White}}]{Fabian1989}
{Fabian}, A.~C., {Rees}, M.~J., {Stella}, L., \& {White}, N.~E. 1989, \mnras,
  238, 729, \dodoi{10.1093/mnras/238.3.729}

\bibitem[{{Fabiani} {et~al.}(2023){Fabiani}, {Capitanio}, {Iaria}, {Poutanen},
  {Gnarini}, {Ursini}, {Farinelli}, {Bobrikova}, {Steiner}, {Svoboda},
  {Anitra}, {Baglio}, {Carotenuto}, {Del Santo}, {Ferrigno}, {Lewis},
  {Russell}, {Russell}, {van den Eijnden}, {Cocchi}, {Di Marco}, {La Monaca},
  {Liu}, {Rankin}, {Weisskopf}, {Xie}, {Bianchi}, {Burderi}, {Di Salvo},
  {Egron}, {Illiano}, {Kaaret}, {Matt}, {Miku{\v{s}}incov{\'a}}, {Muleri},
  {Papitto}, {Agudo}, {Antonelli}, {Bachetti}, {Baldini}, {Baumgartner},
  {Bellazzini}, {Bongiorno}, {Bonino}, {Brez}, {Bucciantini}, {Castellano},
  {Cavazzuti}, {Chen}, {Ciprini}, {Costa}, {De Rosa}, {Del Monte}, {Di Gesu},
  {Di Lalla}, {Donnarumma}, {Doroshenko}, {Dov{\v{c}}iak}, {Ehlert}, {Enoto},
  {Evangelista}, {Ferrazzoli}, {Garcia}, {Gunji}, {Hayashida}, {Heyl},
  {Iwakiri}, {Jorstad}, {Karas}, {Kislat}, {Kitaguchi}, {Kolodziejczak},
  {Krawczynski}, {Latronico}, {Liodakis}, {Maldera}, {Manfreda}, {Marin},
  {Marinucci}, {Marscher}, {Marshall}, {Massaro}, {Mitsuishi}, {Mizuno},
  {Negro}, {Ng}, {O'Dell}, {Omodei}, {Oppedisano}, {Pavlov}, {Peirson},
  {Perri}, {Pesce-Rollins}, {Petrucci}, {Pilia}, {Possenti}, {Puccetti},
  {Ramsey}, {Ratheesh}, {Roberts}, {Romani}, {Sgr{\`o}}, {Slane}, {Soffitta},
  {Spandre}, {Swartz}, {Tamagawa}, {Tavecchio}, {Taverna}, {Tawara}, {Tennant},
  {Thomas}, {Tombesi}, {Trois}, {Tsygankov}, {Turolla}, {Vink}, {Wu}, \&
  {Zane}}]{Fabiani23}
{Fabiani}, S., {Capitanio}, F., {Iaria}, R., {et~al.} 2023, \aap, in press,
  arXiv:2310.06788, \dodoi{10.1051/0004-6361/202347374}

\bibitem[{{Farinelli} {et~al.}(2023){Farinelli}, {Fabiani}, {Poutanen},
  {Ursini}, {Ferrigno}, {Bianchi}, {Cocchi}, {Capitanio}, {De Rosa}, {Gnarini},
  {Kislat}, {Matt}, {Mikusincova}, {Muleri}, {Agudo}, {Antonelli}, {Bachetti},
  {Baldini}, {Baumgartner}, {Bellazzini}, {Bongiorno}, {Bonino}, {Brez},
  {Bucciantini}, {Castellano}, {Cavazzuti}, {Ciprini}, {Costa}, {Del Monte},
  {Di Gesu}, {Di Lalla}, {Di Marco}, {Donnarumma}, {Doroshenko},
  {Dov{\v{c}}iak}, {Ehlert}, {Enoto}, {Evangelista}, {Ferrazzoli}, {Garcia},
  {Gunji}, {Hayashida}, {Heyl}, {Iwakiri}, {Jorstad}, {Karas}, {Kitaguchi},
  {Kolodziejczak}, {Krawczynski}, {La Monaca}, {Latronico}, {Liodakis},
  {Maldera}, {Manfreda}, {Marin}, {Marscher}, {Marshall}, {Mitsuishi},
  {Mizuno}, {Ng}, {O'Dell}, {Omodei}, {Oppedisano}, {Papitto}, {Pavlov},
  {Peirson}, {Perri}, {Pesce-Rollins}, {Petrucci}, {Pilia}, {Possenti},
  {Puccetti}, {Ramsey}, {Rankin}, {Ratheesh}, {Romani}, {Sgr{\`o}}, {Slane},
  {Soffitta}, {Spandre}, {Tamagawa}, {Tavecchio}, {Taverna}, {Tawara},
  {Tennant}, {Thomas}, {Tombesi}, {Trois}, {Tsygankov}, {Turolla}, {Vink},
  {Weisskopf}, {Wu}, {Xie}, \& {Zane}}]{Farinelli23}
{Farinelli}, R., {Fabiani}, S., {Poutanen}, J., {et~al.} 2023, \mnras, 519,
  3681, \dodoi{10.1093/mnras/stac3726}

\bibitem[{{Ferrazzoli} {et~al.}(2020){Ferrazzoli}, {Muleri}, {Lefevre},
  {Morbidini}, {Amici}, {Brienza}, {Costa}, {Del Monte}, {Di Marco}, {Di
  Persio}, {Donnarumma}, {Fabiani}, {La Monaca}, {Loffredo}, {Maiolo}, {Maita},
  {Piazzolla}, {Ramsey}, {Rankin}, {Ratheesh}, {Rubini}, {Sarra}, {Soffitta},
  {Tobia}, \& {Xie}}]{Ferrazzoli20}
{Ferrazzoli}, R., {Muleri}, F., {Lefevre}, C., {et~al.} 2020, JATIS, 6, 048002,
  \dodoi{10.1117/1.JATIS.6.4.048002}

\bibitem[{{Fomalont} {et~al.}(2001{\natexlab{a}}){Fomalont}, {Geldzahler}, \&
  {Bradshaw}}]{Fomalont01}
{Fomalont}, E.~B., {Geldzahler}, B.~J., \& {Bradshaw}, C.~F.
  2001{\natexlab{a}}, \apjl, 553, L27, \dodoi{10.1086/320490}

\bibitem[{{Fomalont} {et~al.}(2001{\natexlab{b}}){Fomalont}, {Geldzahler}, \&
  {Bradshaw}}]{Fomalont2001b}
---. 2001{\natexlab{b}}, \apj, 558, 283, \dodoi{10.1086/322479}

\bibitem[{{Galloway} {et~al.}(2014){Galloway}, {Premachandra}, {Steeghs},
  {Marsh}, {Casares}, \& {Cornelisse}}]{Galloway14}
{Galloway}, D.~K., {Premachandra}, S., {Steeghs}, D., {et~al.} 2014, \apj, 781,
  14, \dodoi{10.1088/0004-637X/781/1/14}

\bibitem[{{Garc{\'\i}a} {et~al.}(2022){Garc{\'\i}a}, {Dauser}, {Ludlam},
  {Parker}, {Fabian}, {Harrison}, \& {Wilms}}]{Garcia22}
{Garc{\'\i}a}, J.~A., {Dauser}, T., {Ludlam}, R., {et~al.} 2022, \apj, 926, 13,
  \dodoi{10.3847/1538-4357/ac3cb7}

\bibitem[{{Gendreau} {et~al.}(2016){Gendreau}, {Arzoumanian}, {Adkins},
  {Albert}, {Anders}, {Aylward}, {Baker}, {Balsamo}, {Bamford}, {Benegalrao},
  {Berry}, {Bhalwani}, {Black}, {Blaurock}, {Bronke}, {Brown}, {Budinoff},
  {Cantwell}, {Cazeau}, {Chen}, {Clement}, {Colangelo}, {Coleman},
  {Coopersmith}, {Dehaven}, {Doty}, {Egan}, {Enoto}, {Fan}, {Ferro}, {Foster},
  {Galassi}, {Gallo}, {Green}, {Grosh}, {Ha}, {Hasouneh}, {Heefner}, {Hestnes},
  {Hoge}, {Jacobs}, {J{\o}rgensen}, {Kaiser}, {Kellogg}, {Kenyon}, {Koenecke},
  {Kozon}, {LaMarr}, {Lambertson}, {Larson}, {Lentine}, {Lewis}, {Lilly},
  {Liu}, {Malonis}, {Manthripragada}, {Markwardt}, {Matonak}, {Mcginnis},
  {Miller}, {Mitchell}, {Mitchell}, {Mohammed}, {Monroe}, {Montt de Garcia},
  {Mul{\'e}}, {Nagao}, {Ngo}, {Norris}, {Norwood}, {Novotka}, {Okajima},
  {Olsen}, {Onyeachu}, {Orosco}, {Peterson}, {Pevear}, {Pham}, {Pollard},
  {Pope}, {Powers}, {Powers}, {Price}, {Prigozhin}, {Ramirez}, {Reid},
  {Remillard}, {Rogstad}, {Rosecrans}, {Rowe}, {Sager}, {Sanders}, {Savadkin},
  {Saylor}, {Schaeffer}, {Schweiss}, {Semper}, {Serlemitsos}, {Shackelford},
  {Soong}, {Struebel}, {Vezie}, {Villasenor}, {Winternitz}, {Wofford},
  {Wright}, {Yang}, \& {Yu}}]{Gendreau16}
{Gendreau}, K.~C., {Arzoumanian}, Z., {Adkins}, P.~W., {et~al.} 2016, in
  \procspie, Vol. 9905, Space Telescopes and Instrumentation 2016: Ultraviolet
  to Gamma Ray, ed. J.-W.~A. {den Herder}, T.~{Takahashi}, \& M.~{Bautz},
  99051H, \dodoi{10.1117/12.2231304}

\bibitem[{{Giacconi} {et~al.}(1962){Giacconi}, {Gursky}, {Paolini}, \&
  {Rossi}}]{Giacconi62}
{Giacconi}, R., {Gursky}, H., {Paolini}, F.~R., \& {Rossi}, B.~B. 1962, \prl,
  9, 439, \dodoi{10.1103/PhysRevLett.9.439}

\bibitem[{{Gilfanov} {et~al.}(2003){Gilfanov}, {Revnivtsev}, \&
  {Molkov}}]{gilfanov2003}
{Gilfanov}, M., {Revnivtsev}, M., \& {Molkov}, S. 2003, \aap, 410, 217,
  \dodoi{10.1051/0004-6361:20031141}

\bibitem[{{Gnarini} {et~al.}(2022){Gnarini}, {Ursini}, {Matt}, {Bianchi},
  {Capitanio}, {Cocchi}, {Farinelli}, \& {Zhang}}]{gnarini2022}
{Gnarini}, A., {Ursini}, F., {Matt}, G., {et~al.} 2022, \mnras, 514, 2561,
  \dodoi{10.1093/mnras/stac1523}

\bibitem[{{Gottlieb} {et~al.}(1975){Gottlieb}, {Wright}, \&
  {Liller}}]{Gottlieb75}
{Gottlieb}, E.~W., {Wright}, E.~L., \& {Liller}, W. 1975, \apjl, 195, L33,
  \dodoi{10.1086/181703}

\bibitem[{{Grefenstette} {et~al.}(2022){Grefenstette}, {Brightman}, {Earnshaw},
  {Forster}, {Madsen}, \& {Miyasaka}}]{Grefenstette22}
{Grefenstette}, B., {Brightman}, M., {Earnshaw}, H.~P., {et~al.} 2022, arXiv
  e-prints, arXiv:2206.04058, \dodoi{10.48550/arXiv.2206.04058}

\bibitem[{{Haardt} \& {Matt}(1993)}]{Haardt93}
{Haardt}, F., \& {Matt}, G. 1993, \mnras, 261, 346,
  \dodoi{10.1093/mnras/261.2.346}

\bibitem[{{Harrison} {et~al.}(2013){Harrison}, {Craig}, {Christensen},
  {Hailey}, {Zhang}, {Boggs}, {Stern}, {Cook}, {Forster}, {Giommi},
  {Grefenstette}, {Kim}, {Kitaguchi}, {Koglin}, {Madsen}, {Mao}, {Miyasaka},
  {Mori}, {Perri}, {Pivovaroff}, {Puccetti}, {Rana}, {Westergaard}, {Willis},
  {Zoglauer}, {An}, {Bachetti}, {Barri{\`e}re}, {Bellm}, {Bhalerao},
  {Brejnholt}, {Fuerst}, {Liebe}, {Markwardt}, {Nynka}, {Vogel}, {Walton},
  {Wik}, {Alexander}, {Cominsky}, {Hornschemeier}, {Hornstrup}, {Kaspi},
  {Madejski}, {Matt}, {Molendi}, {Smith}, {Tomsick}, {Ajello}, {Ballantyne},
  {Balokovi{\'c}}, {Barret}, {Bauer}, {Blandford}, {Brandt}, {Brenneman},
  {Chiang}, {Chakrabarty}, {Chenevez}, {Comastri}, {Dufour}, {Elvis}, {Fabian},
  {Farrah}, {Fryer}, {Gotthelf}, {Grindlay}, {Helfand}, {Krivonos}, {Meier},
  {Miller}, {Natalucci}, {Ogle}, {Ofek}, {Ptak}, {Reynolds}, {Rigby},
  {Tagliaferri}, {Thorsett}, {Treister}, \& {Urry}}]{Harrison2013}
{Harrison}, F.~A., {Craig}, W.~W., {Christensen}, F.~E., {et~al.} 2013, \apj,
  770, 103, \dodoi{10.1088/0004-637X/770/2/103}

\bibitem[{{Hasinger} \& {van der Klis}(1989)}]{hasinger89}
{Hasinger}, G., \& {van der Klis}, M. 1989, \aap, 225, 79

\bibitem[{{HI4PI Collaboration} {et~al.}(2016){HI4PI Collaboration}, {Ben
  Bekhti}, {Fl{\"o}er}, {Keller}, {Kerp}, {Lenz}, {Winkel}, {Bailin},
  {Calabretta}, {Dedes}, {Ford}, {Gibson}, {Haud}, {Janowiecki}, {Kalberla},
  {Lockman}, {McClure-Griffiths}, {Murphy}, {Nakanishi}, {Pisano}, \&
  {Staveley-Smith}}]{HI4PI}
{HI4PI Collaboration}, {Ben Bekhti}, N., {Fl{\"o}er}, L., {et~al.} 2016, \aap,
  594, A116, \dodoi{10.1051/0004-6361/201629178}

\bibitem[{Huppenkothen {et~al.}(2019)Huppenkothen, Bachetti, Stevens, Migliari,
  Balm, Hammad, Khan, Mishra, Rashid, Sharma, Ribeiro, \&
  Blanco}]{huppenkothen2019}
Huppenkothen, D., Bachetti, M., Stevens, A.~L., {et~al.} 2019, ApJ, 881, 39,
  \dodoi{10.3847/1538-4357/ab258d}

\bibitem[{{Inogamov} \& {Sunyaev}(1999)}]{inogamov1999}
{Inogamov}, N.~A., \& {Sunyaev}, R.~A. 1999, Astronomy Letters, 25, 269.
\newblock \doarXiv{astro-ph/9904333}

\bibitem[{{Jayasurya} {et~al.}(2023){Jayasurya}, {Agrawal}, \&
  {Chatterjee}}]{Jayasurya2023}
{Jayasurya}, K.~M., {Agrawal}, V.~K., \& {Chatterjee}, R. 2023, \mnras, 525,
  4657, \dodoi{10.1093/mnras/stad2601}

\bibitem[{{Kalberla} {et~al.}(2005){Kalberla}, {Burton}, {Hartmann}, {Arnal},
  {Bajaja}, {Morras}, \& {P{\"o}ppel}}]{Kalberla05}
{Kalberla}, P.~M.~W., {Burton}, W.~B., {Hartmann}, D., {et~al.} 2005, \aap,
  440, 775, \dodoi{10.1051/0004-6361:20041864}

\bibitem[{{Kislat} {et~al.}(2015){Kislat}, {Clark}, {Beilicke}, \&
  {Krawczynski}}]{Kislat2015}
{Kislat}, F., {Clark}, B., {Beilicke}, M., \& {Krawczynski}, H. 2015,
  Astroparticle Physics, 68, 45, \dodoi{10.1016/j.astropartphys.2015.02.007}

\bibitem[{{Krawczynski} {et~al.}(2022){Krawczynski}, {Muleri}, {Dov{\v{c}}iak},
  {Veledina}, {Rodriguez Cavero}, {Svoboda}, {Ingram}, {Matt}, {Garcia},
  {Loktev}, {Negro}, {Poutanen}, {Kitaguchi}, {Podgorn{\'y}}, {Rankin},
  {Zhang}, {Berdyugin}, {Berdyugina}, {Bianchi}, {Blinov}, {Capitanio}, {Di
  Lalla}, {Draghis}, {Fabiani}, {Kagitani}, {Kravtsov}, {Kiehlmann},
  {Latronico}, {Lutovinov}, {Mandarakas}, {Marin}, {Marinucci}, {Miller},
  {Mizuno}, {Molkov}, {Omodei}, {Petrucci}, {Ratheesh}, {Sakanoi}, {Semena},
  {Skalidis}, {Soffitta}, {Tennant}, {Thalhammer}, {Tombesi}, {Weisskopf},
  {Wilms}, {Zhang}, {Agudo}, {Antonelli}, {Bachetti}, {Baldini}, {Baumgartner},
  {Bellazzini}, {Bongiorno}, {Bonino}, {Brez}, {Bucciantini}, {Castellano},
  {Cavazzuti}, {Ciprini}, {Costa}, {De Rosa}, {Del Monte}, {Di Gesu}, {Di
  Marco}, {Donnarumma}, {Doroshenko}, {Ehlert}, {Enoto}, {Evangelista},
  {Ferrazzoli}, {Gunji}, {Hayashida}, {Heyl}, {Iwakiri}, {Jorstad}, {Karas},
  {Kolodziejczak}, {La Monaca}, {Liodakis}, {Maldera}, {Manfreda}, {Marscher},
  {Marshall}, {Mitsuishi}, {Ng}, {O{\textquoteright}Dell}, {Oppedisano},
  {Papitto}, {Pavlov}, {Peirson}, {Perri}, {Pesce-Rollins}, {Pilia},
  {Possenti}, {Puccetti}, {Ramsey}, {Romani}, {Sgr{\`o}}, {Slane}, {Spandre},
  {Tamagawa}, {Tavecchio}, {Taverna}, {Tawara}, {Thomas}, {Trois}, {Tsygankov},
  {Turolla}, {Vink}, {Wu}, {Xie}, \& {Zane}}]{Krawczynski2022}
{Krawczynski}, H., {Muleri}, F., {Dov{\v{c}}iak}, M., {et~al.} 2022, Science,
  378, 650, \dodoi{10.1126/science.add5399}

\bibitem[{{Kuulkers} {et~al.}(1994){Kuulkers}, {van der Klis}, {Oosterbroek},
  {Asai}, {Dotani}, {van Paradijs}, \& {Lewin}}]{Kuulkers94}
{Kuulkers}, E., {van der Klis}, M., {Oosterbroek}, T., {et~al.} 1994, \aap,
  289, 795

\bibitem[{{Kuulkers} {et~al.}(1997){Kuulkers}, {van der Klis}, {Oosterbroek},
  {van Paradijs}, \& {Lewin}}]{kuulkers97}
{Kuulkers}, E., {van der Klis}, M., {Oosterbroek}, T., {van Paradijs}, J., \&
  {Lewin}, W.~H.~G. 1997, \mnras, 287, 495, \dodoi{10.1093/mnras/287.3.495}

\bibitem[{{Lapidus} \& {Sunyaev}(1985)}]{Lapidus85}
{Lapidus}, I.~I., \& {Sunyaev}, R.~A. 1985, \mnras, 217, 291,
  \dodoi{10.1093/mnras/217.2.291}

\bibitem[{{Leahy} {et~al.}(1983){Leahy}, {Darbro}, {Elsner}, {Weisskopf},
  {Sutherland}, {Kahn}, \& {Grindlay}}]{Leahy83}
{Leahy}, D.~A., {Darbro}, W., {Elsner}, R.~F., {et~al.} 1983, \apj, 266, 160,
  \dodoi{10.1086/160766}

\bibitem[{{Li} {et~al.}(2020){Li}, {Li}, {Tan}, {Yang}, {Ge}, {Zhang}, {Tuo},
  {Wu}, {Liao}, {Zhang}, {Song}, {Zhang}, {Qu}, {Zhang}, {Lu}, {Xu}, {Liu},
  {Cao}, {Chen}, {Nie}, {Zhao}, \& {Li}}]{Li20}
{Li}, X., {Li}, X., {Tan}, Y., {et~al.} 2020, Journal of High Energy
  Astrophysics, 27, 64, \dodoi{10.1016/j.jheap.2020.02.009}

\bibitem[{{Long} {et~al.}(1979){Long}, {Chanan}, {Ku}, \& {Novick}}]{Long1979}
{Long}, K.~S., {Chanan}, G.~A., {Ku}, W.~H.~M., \& {Novick}, R. 1979, \apjl,
  232, L107, \dodoi{10.1086/183045}

\bibitem[{{Long} {et~al.}(2022){Long}, {Feng}, {Li}, {Zhu}, {Wu}, {Huang},
  {Minuti}, {Jiang}, {Yang}, {Citraro}, {Nasimi}, {Yu}, {Jin}, {Zeng}, {An},
  {Jiang}, {Costa}, {Baldini}, {Bellazzini}, {Brez}, {Latronico}, {Sgr{\`o}},
  {Spandre}, {Pinchera}, {Muleri}, \& {Soffitta}}]{Long2022}
{Long}, X., {Feng}, H., {Li}, H., {et~al.} 2022, \apjl, 924, L13,
  \dodoi{10.3847/2041-8213/ac4673}

\bibitem[{{Loskutov} \& {Sobolev}(1982)}]{Loskutov82}
{Loskutov}, V.~M., \& {Sobolev}, V.~V. 1982, Astrofizika, 18, 81

\bibitem[{{Ludlam} {et~al.}(2018){Ludlam}, {Miller}, {Arzoumanian}, {Bult},
  {Cackett}, {Chakrabarty}, {Dauser}, {Enoto}, {Fabian}, {Garc{\'\i}a},
  {Gendreau}, {Guillot}, {Homan}, {Jaisawal}, {Keek}, {La Marr}, {Malacaria},
  {Markwardt}, {Steiner}, \& {Strohmayer}}]{Ludlam18}
{Ludlam}, R.~M., {Miller}, J.~M., {Arzoumanian}, Z., {et~al.} 2018, \apjl, 858,
  L5, \dodoi{10.3847/2041-8213/aabee6}

\bibitem[{{Ludlam} {et~al.}(2022){Ludlam}, {Cackett}, {Garc{\'\i}a}, {Miller},
  {Stevens}, {Fabian}, {Homan}, {Ng}, {Guillot}, {Buisson}, \&
  {Chakrabarty}}]{Ludlam22}
{Ludlam}, R.~M., {Cackett}, E.~M., {Garc{\'\i}a}, J.~A., {et~al.} 2022, \apj,
  927, 112, \dodoi{10.3847/1538-4357/ac5028}

\bibitem[{{Madsen} {et~al.}(2022){Madsen}, {Forster}, {Grefenstette},
  {Harrison}, \& {Miyasaka}}]{madsen22}
{Madsen}, K.~K., {Forster}, K., {Grefenstette}, B., {Harrison}, F.~A., \&
  {Miyasaka}, H. 2022, JATIS, 8, 034003, \dodoi{10.1117/1.JATIS.8.3.034003}

\bibitem[{{Matt}(1993)}]{Matt93}
{Matt}, G. 1993, \mnras, 260, 663, \dodoi{10.1093/mnras/260.3.663}

\bibitem[{{Matt}(2006)}]{Matt06}
---. 2006, AN, 327, 949, \dodoi{10.1002/asna.200610670}

\bibitem[{{Mazzola} {et~al.}(2021){Mazzola}, {Iaria}, {Di Salvo}, {Sanna},
  {Gambino}, {Marino}, {Bozzo}, {Ferrigno}, {Riggio}, {Anitra}, \&
  {Burderi}}]{Mazzola21}
{Mazzola}, S.~M., {Iaria}, R., {Di Salvo}, T., {et~al.} 2021, \aap, 654, A102,
  \dodoi{10.1051/0004-6361/202039983}

\bibitem[{{Miller-Jones} {et~al.}(2019){Miller-Jones}, {Tetarenko}, {Sivakoff},
  {Middleton}, {Altamirano}, {Anderson}, {Belloni}, {Fender}, {Jonker},
  {K{\"o}rding}, {Krimm}, {Maitra}, {Markoff}, {Migliari}, {Mooley}, {Rupen},
  {Russell}, {Russell}, {Sarazin}, {Soria}, \& {Tudose}}]{Miller-Jones2019}
{Miller-Jones}, J. C.~A., {Tetarenko}, A.~J., {Sivakoff}, G.~R., {et~al.} 2019,
  Nature, 569, 374, \dodoi{10.1038/s41586-019-1152-0}

\bibitem[{{Mondal} {et~al.}(2020){Mondal}, {Dewangan}, \&
  {Raychaudhuri}}]{Mondal20}
{Mondal}, A.~S., {Dewangan}, G.~C., \& {Raychaudhuri}, B. 2020, \mnras, 494,
  3177, \dodoi{10.1093/mnras/staa1001}

\bibitem[{{Mondal} {et~al.}(2022){Mondal}, {Raychaudhuri}, {Dewangan}, \&
  {Beri}}]{Mondal22}
{Mondal}, A.~S., {Raychaudhuri}, B., {Dewangan}, G.~C., \& {Beri}, A. 2022,
  \mnras, 516, 1256, \dodoi{10.1093/mnras/stac2321}

\bibitem[{{Motta} \& {Fender}(2019)}]{Motta19}
{Motta}, S.~E., \& {Fender}, R.~P. 2019, \mnras, 483, 3686,
  \dodoi{10.1093/mnras/sty3331}

\bibitem[{{Popham} \& {Sunyaev}(2001)}]{Popham01}
{Popham}, R., \& {Sunyaev}, R. 2001, \apj, 547, 355, \dodoi{10.1086/318336}

\bibitem[{{Poutanen} {et~al.}(1996){Poutanen}, {Nagendra}, \&
  {Svensson}}]{Poutanen96}
{Poutanen}, J., {Nagendra}, K.~N., \& {Svensson}, R. 1996, \mnras, 283, 892,
  \dodoi{10.1093/mnras/283.3.892}

\bibitem[{{Poutanen} \& {Svensson}(1996)}]{PS96}
{Poutanen}, J., \& {Svensson}, R. 1996, \apj, 470, 249, \dodoi{10.1086/177865}

\bibitem[{{Poutanen} {et~al.}(2023){Poutanen}, {Veledina}, \&
  {Beloborodov}}]{Poutanen_2023}
{Poutanen}, J., {Veledina}, A., \& {Beloborodov}, A.~M. 2023, \apjl, 949, L10,
  \dodoi{10.3847/2041-8213/acd33e}

\bibitem[{{Poutanen} \& {Vilhu}(1993)}]{Poutanen93}
{Poutanen}, J., \& {Vilhu}, O. 1993, \aap, 275, 337

\bibitem[{{Psaltis} {et~al.}(1995){Psaltis}, {Lamb}, \& {Miller}}]{Psaltis95}
{Psaltis}, D., {Lamb}, F.~K., \& {Miller}, G.~S. 1995, \apjl, 454, L137,
  \dodoi{10.1086/309780}

\bibitem[{{Rankin} {et~al.}(2023){Rankin}, {La Monaca}, {Di Marco}, {Poutanen},
  {Bobrikova}, {Kravtsov}, {Muleri}, {Pilia}, {Veledina}, {Fender}, {Kaaret},
  {Kim}, {Marinucci}, {Marshall}, {Papitto}, {Tennant}, {Tsygankov},
  {Weisskopf}, {Wu}, {Zane}, {Ambrosino}, {Farinelli}, {Gnarini}, {Agudo},
  {Antonelli}, {Bachetti}, {Baldini}, {Baumgartner}, {Bellazzini}, {Bianchi},
  {Bongiorno}, {Bonino}, {Brez}, {Bucciantini}, {Capitanio}, {Castellano},
  {Cavazzuti}, {Chen}, {Ciprini}, {Costa}, {De Rosa}, {Del Monte}, {Di Gesu},
  {di Lalla}, {Donnarumma}, {Doroshenko}, {Dov{\v{c}}iak}, {Ehlert}, {Enoto},
  {Evangelista}, {Fabiani}, {Ferrazzoli}, {Garcia}, {Gunji}, {Hayashida},
  {Heyl}, {Iwakiri}, {Jorstad}, {Karas}, {Kislat}, {Kitaguchi},
  {Kolodziejczak}, {Krawczynski}, {Latronico}, {Liodakis}, {Maldera},
  {Manfreda}, {Marin}, {Marscher}, {Massaro}, {Matt}, {Mitsuishi}, {Mizuno},
  {Negro}, {Ng}, {O'Dell}, {Omodei}, {Oppedisano}, {Pavlov}, {Peirson},
  {Perri}, {Pesce-Rollins}, {Petrucci}, {Possenti}, {Puccetti}, {Ramsey},
  {Ratheesh}, {Roberts}, {Romani}, {Sgr{\`o}}, {Slane}, {Soffitta}, {Spandre},
  {Swartz}, {Tamagawa}, {Tavecchio}, {Taverna}, {Tawara}, {Thomas}, {Tombesi},
  {Trois}, {Turolla}, {Vink}, \& {Xie}}]{Rankin23}
{Rankin}, J., {La Monaca}, F., {Di Marco}, A., {et~al.} 2023, \apjl, submitted,
  arXiv:2311.04632, \dodoi{10.48550/arXiv.2311.04632}

\bibitem[{Rankin {et~al.}(2023)Rankin, Muleri, Ferrazzoli, Baldini,
  Baumgartner, Costa, Monte, Marco, Dietz, Ehlert, Fabiani, Kaaret, Kim,
  Kolodziejczak, Monaca, Latronico, Lefevre, Manfreda, O'Dell, Perri, Puccetti,
  Ramsey, Ratheesh, Rubini, Sgr{\`o}, Soffitta, Swartz, Tennant, Trois, \&
  Weisskopf}]{Rankin23monitoring}
Rankin, J., Muleri, F., Ferrazzoli, R., {et~al.} 2023, in \procspie, Vol.
  12678, UV, X-Ray, and Gamma-Ray Space Instrumentation for Astronomy XXIII,
  ed. O.~H. Siegmund \& K.~Hoadley, 126780D, \dodoi{10.1117/12.2677264}

\bibitem[{{Revnivtsev} \& {Gilfanov}(2006)}]{Revnivtsev06}
{Revnivtsev}, M.~G., \& {Gilfanov}, M.~R. 2006, \aap, 453, 253,
  \dodoi{10.1051/0004-6361:20053964}

\bibitem[{{Revnivtsev} {et~al.}(2013){Revnivtsev}, {Suleimanov}, \&
  {Poutanen}}]{Revnivtsev13}
{Revnivtsev}, M.~G., {Suleimanov}, V.~F., \& {Poutanen}, J. 2013, \mnras, 434,
  2355, \dodoi{10.1093/mnras/stt1179}

\bibitem[{{Schnittman} \& {Krolik}(2010)}]{Schnittman2010}
{Schnittman}, J.~D., \& {Krolik}, J.~H. 2010, \apj, 712, 908,
  \dodoi{10.1088/0004-637X/712/2/908}

\bibitem[{{Schulz} \& {Wijers}(1993)}]{Schulz93}
{Schulz}, N.~S., \& {Wijers}, R.~A.~M.~J. 1993, \aap, 273, 123

\bibitem[{{Shakura} \& {Sunyaev}(1988)}]{Shakura88}
{Shakura}, N.~I., \& {Sunyaev}, R.~A. 1988, Advances in Space Research, 8, 135,
  \dodoi{10.1016/0273-1177(88)90396-1}

\bibitem[{{Sobolev}(1963)}]{Sobolev63}
{Sobolev}, V.~V. 1963, {A Treatise on Radiative Transfer} (Princeton: Van
  Nostrand)

\bibitem[{Soffitta {et~al.}(2021)Soffitta, Baldini, Bellazzini, Costa,
  Latronico, Muleri, Monte, Fabiani, Minuti, Pinchera, Sgro’, Spandre, Trois,
  Amici, Andersson, Attina’, Bachetti, Barbanera, Borotto, Brez, Brienza,
  Caporale, Cardelli, Carpentiero, Castellano, Castronuovo, Cavalli, Cavazzuti,
  Ceccanti, Centrone, Ciprini, Citraro, D’Amico, D’Alba, Cosimo, Lalla,
  Marco, Persio, Donnarumma, Evangelista, Ferrazzoli, Hayato, Kitaguchi,
  Monaca, Lefevre, Loffredo, Lorenzi, Lucchesi, Magazzu, Maldera, Manfreda,
  Mangraviti, Marengo, Matt, Mereu, Morbidini, Mosti, Nakano, Nasimi, Negri,
  Nenonen, Nuti, Orsini, Perri, Pesce-Rollins, Piazzolla, Pilia, Profeti,
  Puccetti, Rankin, Ratheesh, Rubini, Santoli, Sarra, Scalise, Sciortino,
  Tamagawa, Tardiola, Tobia, Vimercati, \& Xie}]{Soffitta21}
Soffitta, P., Baldini, L., Bellazzini, R., {et~al.} 2021, \aj, 162, 208,
  \dodoi{10.3847/1538-3881/ac19b0}

\bibitem[{{Steeghs} \& {Casares}(2002)}]{Steeghs02}
{Steeghs}, D., \& {Casares}, J. 2002, \apj, 568, 273, \dodoi{10.1086/339224}

\bibitem[{{Stella} \& {Vietri}(1998)}]{Stella98}
{Stella}, L., \& {Vietri}, M. 1998, \apjl, 492, L59, \dodoi{10.1086/311075}

\bibitem[{{Suleimanov} \& {Poutanen}(2006)}]{suleimanov2006}
{Suleimanov}, V., \& {Poutanen}, J. 2006, \mnras, 369, 2036,
  \dodoi{10.1111/j.1365-2966.2006.10454.x}

\bibitem[{{Sunyaev} \& {Titarchuk}(1985)}]{st85}
{Sunyaev}, R.~A., \& {Titarchuk}, L.~G. 1985, \aap, 143, 374

\bibitem[{{Titarchuk} {et~al.}(2014){Titarchuk}, {Seifina}, \&
  {Shrader}}]{Titarchuk14}
{Titarchuk}, L., {Seifina}, E., \& {Shrader}, C. 2014, \apj, 789, 98,
  \dodoi{10.1088/0004-637X/789/2/98}

\bibitem[{{Tomsick} {et~al.}(2018){Tomsick}, {Parker}, {Garc{\'\i}a},
  {Yamaoka}, {Barret}, {Chiu}, {Clavel}, {Fabian}, {F{\"u}rst}, {Gandhi},
  {Grinberg}, {Miller}, {Pottschmidt}, \& {Walton}}]{Tomsick18}
{Tomsick}, J.~A., {Parker}, M.~L., {Garc{\'\i}a}, J.~A., {et~al.} 2018, \apj,
  855, 3, \dodoi{10.3847/1538-4357/aaaab1}

\bibitem[{{Ursini} {et~al.}(2023){Ursini}, {Farinelli}, {Gnarini}, {Poutanen},
  {Bianchi}, {Capitanio}, {Di Marco}, {Fabiani}, {La Monaca}, {Malacaria},
  {Matt}, {Miku{\v{s}}incov{\'a}}, {Cocchi}, {Kaaret}, {Kajava}, {Pilia},
  {Zhang}, {Agudo}, {Antonelli}, {Bachetti}, {Baldini}, {Baumgartner},
  {Bellazzini}, {Bongiorno}, {Bonino}, {Brez}, {Bucciantini}, {Castellano},
  {Cavazzuti}, {Chen}, {Ciprini}, {Costa}, {De Rosa}, {Del Monte}, {Di Gesu},
  {Di Lalla}, {Donnarumma}, {Doroshenko}, {Dov{\v{c}}iak}, {Ehlert}, {Enoto},
  {Evangelista}, {Ferrazzoli}, {Garcia}, {Gunji}, {Hayashida}, {Heyl},
  {Iwakiri}, {Jorstad}, {Karas}, {Kislat}, {Kitaguchi}, {Kolodziejczak},
  {Krawczynski}, {Latronico}, {Liodakis}, {Maldera}, {Manfreda}, {Marin},
  {Marinucci}, {Marscher}, {Marshall}, {Massaro}, {Mitsuishi}, {Mizuno},
  {Muleri}, {Negro}, {Ng}, {O'Dell}, {Omodei}, {Oppedisano}, {Papitto},
  {Pavlov}, {Peirson}, {Perri}, {Pesce-Rollins}, {Petrucci}, {Pilia},
  {Possenti}, {Puccetti}, {Ramsey}, {Rankin}, {Ratheesh}, {Roberts}, {Romani},
  {Sgr{\`o}}, {Slane}, {Soffitta}, {Spandre}, {Swartz}, {Tamagawa},
  {Tavecchio}, {Taverna}, {Tawara}, {Tennant}, {Thomas}, {Tombesi}, {Trois},
  {Tsygankov}, {Turolla}, {Vink}, {Weisskopf}, {Wu}, {Xie}, \&
  {Zane}}]{Ursini23}
{Ursini}, F., {Farinelli}, R., {Gnarini}, A., {et~al.} 2023, \aap, 676, A20,
  \dodoi{10.1051/0004-6361/202346541}

\bibitem[{{van der Klis}(1989)}]{vanderklis89}
{van der Klis}, M. 1989, \araa, 27, 517,
  \dodoi{10.1146/annurev.aa.27.090189.002505}

\bibitem[{{Veledina} {et~al.}(2023){Veledina}, {Muleri}, {Dov{\v{c}}iak},
  {Poutanen}, {Ratheesh}, {Capitanio}, {Matt}, {Soffitta}, {Tennant}, {Negro},
  {Kaaret}, {Costa}, {Ingram}, {Svoboda}, {Krawczynski}, {Bianchi}, {Steiner},
  {Garc{\'\i}a}, {Kravtsov}, {Nitindala}, {Ewing}, {Mastroserio}, {Marinucci},
  {Ursini}, {Tombesi}, {Tsygankov}, {Yang}, {Weisskopf}, {Trushkin}, {Egron},
  {Iacolina}, {Pilia}, {Marra}, {Miku{\v{s}}incov{\'a}}, {Nathan}, {Parra},
  {Petrucci}, {Podgorn{\'y}}, {Tugliani}, {Zane}, {Zhang}, {Agudo},
  {Antonelli}, {Bachetti}, {Baldini}, {Baumgartner}, {Bellazzini}, {Bongiorno},
  {Bonino}, {Brez}, {Bucciantini}, {Castellano}, {Cavazzuti}, {Chen},
  {Ciprini}, {De Rosa}, {Del Monte}, {Di Gesu}, {Di Lalla}, {Di Marco},
  {Donnarumma}, {Doroshenko}, {Ehlert}, {Enoto}, {Evangelista}, {Fabiani},
  {Ferrazzoli}, {Gunji}, {Hayashida}, {Heyl}, {Iwakiri}, {Jorstad}, {Karas},
  {Kislat}, {Kitaguchi}, {Kolodziejczak}, {La Monaca}, {Latronico}, {Liodakis},
  {Maldera}, {Manfreda}, {Marin}, {Marscher}, {Marshall}, {Massaro},
  {Mitsuishi}, {Mizuno}, {Ng}, {O'Dell}, {Omodei}, {Oppedisano}, {Papitto},
  {Pavlov}, {Peirson}, {Perri}, {Pesce-Rollins}, {Possenti}, {Puccetti},
  {Ramsey}, {Rankin}, {Roberts}, {Romani}, {Sgr{\`o}}, {Slane}, {Spandre},
  {Swartz}, {Tamagawa}, {Tavecchio}, {Taverna}, {Tawara}, {Thomas}, {Trois},
  {Turolla}, {Vink}, {Wu}, \& {Xie}}]{Veledina23}
{Veledina}, A., {Muleri}, F., {Dov{\v{c}}iak}, M., {et~al.} 2023, \apjl, 958,
  L16, \dodoi{10.3847/2041-8213/ad0781}

\bibitem[{{Weisskopf} {et~al.}(2022){Weisskopf}, {Soffitta}, {Baldini},
  {Ramsey}, {O'Dell}, {Romani}, {Matt}, {Deininger}, {Baumgartner},
  {Bellazzini}, {Costa}, {Kolodziejczak}, {Latronico}, {Marshall}, {Muleri},
  {Bongiorno}, {Tennant}, {Bucciantini}, {Dovciak}, {Marin}, {Marscher},
  {Poutanen}, {Slane}, {Turolla}, {Kalinowski}, {Di Marco}, {Fabiani},
  {Minuti}, {La Monaca}, {Pinchera}, {Rankin}, {Sgro'}, {Trois}, {Xie},
  {Alexander}, {Allen}, {Amici}, {Andersen}, {Antonelli}, {Antoniak},
  {Attina'}, {Barbanera}, {Bachetti}, {Baggett}, {Bladt}, {Brez}, {Bonino},
  {Boree}, {Borotto}, {Breeding}, {Brienza}, {Bygott}, {Caporale}, {Cardelli},
  {Carpentiero}, {Castellano}, {Castronuovo}, {Cavalli}, {Cavazzuti},
  {Ceccanti}, {Centrone}, {Citraro}, {D' Amico}, {D'Alba}, {Di Gesu}, {Del
  Monte}, {Dietz}, {Di Lalla}, {Di Persio}, {Dolan}, {Donnarumma},
  {Evangelista}, {Ferrant}, {Ferrazzoli}, {Ferrie}, {Footdale}, {Forsyth},
  {Foster}, {Garelick}, {Gunji}, {Gurnee}, {Head}, {Hibbard}, {Johnson},
  {Kelly}, {Kilaru}, {Lefevre}, {Le Roy}, {Loffredo}, {Lorenzi}, {Lucchesi},
  {Maddox}, {Magazzu}, {Maldera}, {Manfreda}, {Mangraviti}, {Marengo},
  {Marrocchesi}, {Massaro}, {Mauger}, {McCracken}, {McEachen}, {Mize}, {Mereu},
  {Mitchell}, {Mitsuishi}, {Morbidini}, {Mosti}, {Nasimi}, {Negri}, {Negro},
  {Nguyen}, {Nitschke}, {Nuti}, {Onizuka}, {Oppedisano}, {Orsini}, {Osborne},
  {Pacheco}, {Paggi}, {Painter}, {Pavelitz}, {Pentz}, {Piazzolla}, {Perri},
  {Pesce-Rollins}, {Peterson}, {Pilia}, {Profeti}, {Puccetti}, {Ranganathan},
  {Ratheesh}, {Reedy}, {Root}, {Rubini}, {Ruswick}, {Sanchez}, {Sarra},
  {Santoli}, {Scalise}, {Sciortino}, {Schroeder}, {Seek}, {Sosdian}, {Spandre},
  {Speegle}, {Tamagawa}, {Tardiola}, {Tobia}, {Thomas}, {Valerie}, {Vimercati},
  {Walden}, {Weddendorf}, {Wedmore}, {Welch}, {Zanetti}, \&
  {Zanetti}}]{Weisskopf2022}
{Weisskopf}, M.~C., {Soffitta}, P., {Baldini}, L., {et~al.} 2022, JATIS, 8,
  026002, \dodoi{10.1117/1.JATIS.8.2.026002}

\bibitem[{{Wilkins}(2018)}]{Wilkins2018}
{Wilkins}, D.~R. 2018, \mnras, 475, 748, \dodoi{10.1093/mnras/stx3167}

\bibitem[{{Wilms} {et~al.}(2000){Wilms}, {Allen}, \&
  {McCray}}]{2000ApJ...542..914W}
{Wilms}, J., {Allen}, A., \& {McCray}, R. 2000, \apj, 542, 914,
  \dodoi{10.1086/317016}

\bibitem[{{Zdziarski} {et~al.}(1996){Zdziarski}, {Johnson}, \&
  {Magdziarz}}]{nthcomp1}
{Zdziarski}, A.~A., {Johnson}, W.~N., \& {Magdziarz}, P. 1996, \mnras, 283,
  193, \dodoi{10.1093/mnras/283.1.193}

\bibitem[{{Zhang} {et~al.}(2014){Zhang}, {Lu}, {Zhang}, \& {Li}}]{Zhang14}
{Zhang}, S., {Lu}, F.~J., {Zhang}, S.~N., \& {Li}, T.~P. 2014, in \procspie,
  Vol. 9144, Space Telescopes and Instrumentation 2014: Ultraviolet to Gamma
  Ray, ed. T.~{Takahashi}, J.-W.~A. {den Herder}, \& M.~{Bautz}, 914421,
  \dodoi{10.1117/12.2054144}

\bibitem[{{Zhang} {et~al.}(2020){Zhang}, {Li}, {Lu}, {Song}, {Xu}, {Liu},
  {Chen}, {Cao}, {Bu}, {Chang}, {Chen}, {Chen}, {Chen}, {Chen}, {Chen}, {Cui},
  {Cui}, {Deng}, {Dong}, {Du}, {Fu}, {Gao}, {Gao}, {Gao}, {Ge}, {Gu}, {Guan},
  {Gungor}, {Guo}, {Han}, {Hu}, {Huang}, {Huo}, {Jia}, {Jiang}, {Jiang}, {Jin},
  {Jin}, {Li}, {Li}, {Li}, {Li}, {Li}, {Li}, {Li}, {Li}, {Li}, {Li}, {Li},
  {Liang}, {Liao}, {Liu}, {Liu}, {Liu}, {Liu}, {Liu}, {Liu}, {Lu}, {Lu}, {Luo},
  {Ma}, {Meng}, {Nang}, {Nie}, {Ou}, {Qu}, {Sai}, {Shang}, {Shen}, {Sun},
  {Tan}, {Tao}, {Tuo}, {Wang}, {Wang}, {Wang}, {Wang}, {Wang}, {Wang}, {Wang},
  {Wen}, {Wu}, {Wu}, {Wu}, {Xiao}, {Xiong}, {Yan}, {Yang}, {Yang}, {Yang},
  {Yi}, {Yuan}, {Zhang}, {Zhang}, {Zhang}, {Zhang}, {Zhang}, {Zhang}, {Zhang},
  {Zhang}, {Zhang}, {Zhang}, {Zhang}, {Zhang}, {Zhang}, {Zhang}, {Zhang},
  {Zhang}, {Zhang}, {Zhang}, {Zhang}, {Zhang}, {Zhao}, {Zhao}, {Zheng}, {Zhou},
  {Zhu}, {Zhu}, {Zhuang}, \& {Insight-HXMT Team}}]{Zhang20}
{Zhang}, S.-N., {Li}, T., {Lu}, F., {et~al.} 2020, Science China Physics,
  Mechanics, and Astronomy, 63, 249502, \dodoi{10.1007/s11433-019-1432-6}

\bibitem[{{{\.Z}ycki} {et~al.}(1999){{\.Z}ycki}, {Done}, \& {Smith}}]{nthcomp2}
{{\.Z}ycki}, P.~T., {Done}, C., \& {Smith}, D.~A. 1999, \mnras, 309, 561,
  \dodoi{10.1046/j.1365-8711.1999.02885.x}

\end{thebibliography}
\bibliographystyle{aasjournal}



\end{document}